\documentclass[12pt,a4paper]{article}
\usepackage{amsmath,amsfonts,amsthm}
\allowdisplaybreaks[3]
\numberwithin{equation}{section}

\usepackage[body={16cm,23.2cm}]{geometry}
\newcommand{\fs}{{\mathbf f}}
\newcommand{\es}{{\mathbf e}}
\newcommand{\hs}{{\mathbf h}}
\newcommand{\Rf}{{\mathbf R}}
\newcommand{\Kf}{{\mathbf K}}
\newcommand{\Lf}{{\mathbf L}}
\newcommand{\Lfc}{\Check{\mathbf L}}
\newcommand{\Lfb}{\overline{\mathbf L}}
\newcommand{\Lfcb}{\Check{\overline{\mathbf L}}}
\newcommand{\Lc}{{\mathcal L}}
\newcommand{\Lcb}{\overline{\mathcal L}}
\newcommand{\Rc}{{\mathcal R}}
\newcommand{\Rcb}{\overline{\mathcal R}}
\newcommand{\Kfcb}{\Check{\overline{\mathbf K}}}
\newcommand{\g}{{\mathbf g}}
\newcommand{\G}{{\mathbf G}}
\newcommand{\Gb}{\overline{\mathbf G}}

\begin{document}
\title{Universal Baxter TQ-relations for open boundary quantum integrable systems
}
\author{Zengo Tsuboi
\\[8pt]
{\sl
Pacific Quantum Center, Far Eastern Federal University, 
}
\\
{\sl
Sukhanova 8, Vladivostok, 690950, Russia
%
%
%
}
} 
\date{}
\maketitle
\begin{abstract}
Based on properties of the universal R-matrix, we 
derive universal Baxter TQ-relations for quantum integrable systems 
with (diagonal) open  boundaries  associated with $U_{q}(\widehat{sl_{2}})$. 
The Baxter TQ-relations for the open XXZ-spin chain are images of 
these universal Baxter TQ-relations. 
\end{abstract}
Keywords: Baxter Q-operator, 
Baxter TQ-relation, 
K-operator, 
L-operator,
reflection equation, 
universal R-matrix
\\[3pt]
Nuclear Physics B 963 (2021) 115286
\\
https://doi.org/10.1016/j.nuclphysb.2020.115286
\section{Introduction}
Baxter Q-operators \cite{Bax72} are fundamental objects in quantum integrable systems. 
They give information about the eigenfunctions and Bethe roots. 
In particular, Bazhanov, Lukyanov and  Zamolodchikov \cite{BLZ97} defined 
Baxter Q-operators as traces of the universal R-matrix over q-oscillator representations of one of the Borel 
subalgebras of the quantum affine algebra $U_{q}(\widehat{sl_{2}})$. 
The universal R-matrix is an element of the tensor product of two Borel subalgebras. 
Hence,  the Baxter Q-operators are universal in the sense that they are elements of a  
Borel subalgebra and thus do not depend on the concrete quantum space of states on which 
the operators act.  Baxter Q-operators for concrete physical models can be obtained 
 by specifying representations of the Borel subalgebra. 
Much work has been done related to 
this `q-oscillator construction' of Baxter Q-operators 
 (see, for example, the following papers and references therein: \cite{AF96,BHK02,KZ05,Korff05,BJMST06,Kojima08,BT08,BGKNR10,Tsuboi12,BGKNR12,
 KT14,Mangazeev14,MT15,NR16,T19-1} for the trigonometric case; \cite{BFLMS10,FLMS10,RT14,Fr20,FFK20} 
  for the rational case;  \cite{PG92,DM06,KLT12,AKLTZ11,PSZ16} for some other methods). 
Moreover, systematic studies related to this from the point of view of the asymptotic representation theory of quantum 
affine algebras were done in \cite{HJ11,FH13,Zhang14}. 

All these works are for quantum integrable systems with periodic boundary condition.
In contrast with the case for models with periodic boundary condition, there are only a few works \cite{FS15,BT18,VW20} on 
the q-oscillator construction
 \footnote{See also \cite{DKM03,DM05,YNZ05,LP14} for different approach.}
 of Baxter Q-operators for models with open boundary conditions. 
 The first breakthrough
  on this topic was brought by 
Frassek and Szecsenyi \cite{FS15} who proposed Baxter Q-operators for the (diagonal) open XXX-spin chain ($q=1$ case). 
In \cite{BT18}, we proposed  universal Baxter Q-operators for 
(diagonal) open boundary quantum integrable systems  associated with $U_{q}(\widehat{sl_{2}})$ and gave 
Baxter Q-operators for open XXZ-spin chains as holomorphic images (for a tensor product 
of the fundamental representation) of them. 
One of the fundamental equations for Baxter Q-operators are the so-called 
Baxter TQ-relations \cite{Bax72}. 
Vlaar and Weston \cite{VW20} proved an operator Baxter  TQ-relation
 \footnote{
 On the level of the eigenvalues, the Baxter  TQ-relation for the open XXZ-spin chain was derived in \cite{Skly} 
 (see also, \cite{KMN14} and references therein). 
 We also remark that a proof of an operator Baxter TQ-relation for 
 the (diagonal)  open XXX-spin chain was previously given in \cite{FS15}.
}
 for the (diagonal)  open XXZ-spin chain based on 
 a representation theoretical method. Thus a Baxter Q-operator for the open XXZ-spin chain proposed in \cite{BT18} 
 indeed satisfies the TQ-relation. 
The purpose of this paper is to supplement \cite{BT18} and give universal analogues of Baxter TQ-relations for (diagonal) open 
boundary quantum integrable systems. Our universal TQ-relations are equations in $U_{q}(\widehat{sl_{2}})$. 
The Baxter TQ-relation in \cite{VW20} follows from our universal Baxter TQ-relation as 
a holomorphic image. 

The layout of this paper is as follows. 
In Section 2, we summarize the definitions of quantum algebras. 
In Section 3, we review L-operators, which are building blocks of T-and Q-operators. 
In particular, the L-operators are defined as holomorphic images of the universal R-matrix.  
In Section 4, we recall solutions  (K-operators)  \cite{BT18} of the reflection equation, 
which are asymptotic limits of solutions  of the 
intertwining relations for the augmented q-Onsager algebra \cite{IT09,BB12}. 
The solutions
are expressed in terms of  Cartan elements of q-oscillator algebras.  
In Section 5, we present the universal Baxter TQ-relations, which are our main results.  
In Appendix A, we review Khoroshkin and Tolstoy's explicit formula \cite{TK92,KT94} of the universal R-matrix. 
In Appendix B, we explain the derivation of the universal TQ-relations. 
In Appendix C, we reconsider the dressed reflection equation in relation to the universal R-matrix. 
In Appendix D, we discuss unitarity relations of R-operators in general situation. 
Throughout the paper we assume that the deformation parameter $q$ is not a root of unity and 
use the following notation. 

\vspace{6pt}

\noindent 
{\bf Notation:} 
\begin{itemize}
\item
For any elements  $X,Y$ of the quantum algebras, the q-commutator is defined by   
$[X,Y]_{q}=XY-qYX$. In particular, we set $[X,Y]_{1}=[X,Y]$.

 %
 \item 
 The q-Pochhammer symbol is defined by 
$(x;q)_{k}=\prod_{j=0}^{k-1}(1-xq^{j})$. In particular, we will use 
$(x;q)_{\infty}=\lim_{k \to \infty}(x;q)_{k}=\prod_{j=0}^{\infty}(1-xq^{j})$ for $|q|<1$. 
For more detail, see for example, page 38 in \cite{KS97}. 
\item
A q-analogue of the exponential function is defined by   
\begin{align}
\exp_{q}(x)=1+\sum_{k=1}^{\infty} \frac{x^{k}}{(k)_{q} !}, 
\label{qexpdef}
\end{align}
 where 
$(k)_{q}! =(1)_{q}(2)_{q}\cdots (k)_{q}$, 
$ (0)_{q}!=1$, $(k)_{q}= (1-q^{k})/(1-q)$. 
This has infinite product expressions.
\begin{align}
\exp_{q}(x)=
\begin{cases} 
((1-q)x;q)^{-1}_{\infty} & \text{for} \quad |q|<1,  \\
((q^{-1}-1)x;q^{-1})_{\infty} & \text{for}\quad  |q|>1.
\end{cases}
 \label{qexp}
\end{align}
Thus the inverse of \eqref{qexpdef} can be obtained via 
$\exp_{q}(x)^{-1}=\exp_{q^{-1}}(-x)$. 
For more detail, see for example, page 47 in \cite{KS97}. 

\item 
We introduce free parameters
  $s_{0},s_{1} \in {\mathbb Z}$. In particular, we set $s=s_{0}+s_{1}$. 
  
\item 
$\lambda=q-q^{-1}$, $[x]_{q}=(q^{x}-q^{-x}) \lambda^{-1}$ for $x \in \mathbb{C}$.
\end{itemize}
 \section{Quantum algebras}
In this section, we review quantum algebras. 
There is overlap among this section and 
 the corresponding sections in \cite{BT18,T19}. 
We also refer to  \cite{CP95,KS97,Ja96} for review on this subject.
\subsection{The quantum affine algebra $U_{q}(\widehat{sl_2})$}
The quantum affine algebra $U_{q}(\widehat{sl_2})$ (at level 0, i.e. the quantum loop algebra)  is a Hopf algebra
generated by the elements
 $e_{i},f_{i},q^{\xi h_{i}}$ for 
$i \in \{0,1 \}$ and $\xi \in {\mathbb C}$ obeying the following relations:
\begin{align}
& q^{0 h_{i}}=q^{0}=1, \quad q^{\xi h_{i}}q^{\eta h_{i}}=q^{(\xi +\eta )h_{i}},  
\quad q^{\xi h_{0}}q^{\xi h_{1}}=1,
\label{sl2h-def1}
\\[6pt]
&[e_{i},f_{j}]=\delta_{ij} \frac{q^{h_{i}} -q^{-h_{i}} }{q-q^{-1}},
\quad q^{\xi h_{i}} e_{j}q^{-\xi h_{i}} =q^{\xi a_{ij} }e_{j}, \quad
q^{\xi h_{i}} f_{j}q^{-\xi h_{i}}  =q^{-\xi a_{ij} }f_{j}, 
\label{sl2h-def2}
\\[6pt]
&[e_{i},[e_{i},[e_{i},e_{j}]_{q^{2}}]]_{q^{-2}}=
[f_{i},[f_{i},[f_{i},f_{j}]_{q^{-2}}]]_{q^{2}}=0
\qquad i \ne j , 
\quad \xi, \eta \in {\mathbb C},
\label{sl2h-def3}
\end{align}
where $(a_{ij})_{0 \le i,j\le 1}$ is the
Cartan matrix
\begin{align}\nonumber
(a_{ij})_{0 \le i,j\le 1}=
\begin{pmatrix}
2& -2 \\
-2 & 2
\end{pmatrix}
.
\end{align}
%
The algebra has an automorphism $\sigma$  defined by 
\begin{align}
\begin{split}
& \sigma (e_{0})=e_{1}, \qquad \sigma (f_{0})=f_{1}, \qquad \sigma (q^{\xi h_{0}})=q^{\xi h_{1}},  
\\[6pt]
& \sigma (e_{1})=e_{0}, \qquad \sigma (f_{1})=f_{0}, \qquad \sigma (q^{\xi h_{1}})=q^{\xi h_{0}}, 
\qquad \sigma(q)=q .
\end{split}
\label{auto1}
\end{align}
The algebra also has an  anti-automorphism $^{t}$  defined by
\footnote{For any $a\in {\mathbb C}$ and a Cartan element ${\mathcal H}$, 
we denote $q^{a}q^{{\mathcal H}}$ as $q^{a+{\mathcal H}}$. } 
\begin{align}
& 
e^{t}_{i}=q^{-1-h_{i}}f_{i}, \quad f^{t}_{i}=e_{i}q^{1+h_{i}}, \quad (q^{\xi h_{i}})^{t}=q^{\xi h_{i}},
\quad q^{t}=q,
\quad i=0,1.
\label{a-auto1}
\end{align}
Note that this means $\sigma(ab)=\sigma(a)\sigma(b)$  and 
 $(ab)^{t}=b^{t} a^{t}$ 
for $a,b \in U_{q}(\widehat{sl_2})$. 
We use the following co-multiplication
  $ \Delta : U_{q}(\widehat{sl_2}) \to U_{q}(\widehat{sl_2}) \otimes U_{q}(\widehat{sl_2})$:
\begin{align}
\Delta (e_{i})&=e_{i} \otimes 1 + q^{-h_{i}} \otimes e_{i}, \nonumber\\
\Delta (f_{i})&=f_{i} \otimes q^{h_{i}} + 1 \otimes f_{i},\label{copro-h} \\
\Delta (q^{\xi h_{i}})&=q^{\xi h_{i}} \otimes q^{\xi h_{i}}. \nonumber
\end{align}
We will also use the opposite co-multiplication defined by
\begin{align}
\Delta'={\mathfrak p}\circ \Delta,\qquad {\mathfrak p}
(X\otimes Y)=
Y\otimes X,\qquad X,Y\in U_{q}(\widehat{sl_{2}}).
\end{align}
Co-unit, anti-pode and grading element $d$ are not  used  in the present 
paper.

The Borel subalgebras ${\mathcal B}_{+}$  and ${\mathcal B}_{-}$ are generated by the elements 
$e_{i}, q^{\xi h_{i}} $ and $f_{i},  q^{\xi h_{i}}$, respectively, where
$i \in \{0,1 \}$, $\xi \in {\mathbb C}$.
There exists a unique element \cite{Dr85,KT92-1} 
${\mathcal R}$ in a completion of $ {\mathcal B}_{+} \otimes {\mathcal B}_{-} $ 
called the universal R-matrix which satisfies the following 
relations
\begin{align}
\begin{split}
\Delta'(a)\ {\mathcal R}&={\mathcal R}\ \Delta(a)
\qquad \text{for} \quad \forall\ a\in U_{q}(\widehat{{sl}_{2}}),   \\[6pt]
(\Delta\otimes 1)\, 
{\mathcal R}&={\mathcal R}_{13}\, {\mathcal R}_{23}, \\[6pt]
(1\otimes \Delta)\, {\mathcal R}&={\mathcal R}_{13}
{\mathcal R}_{12} ,
\end{split}
\label{R-def}
\end{align}
where
\footnote{We will use similar notation for the L-operators 
to indicate the space on which they non-trivially act.} 
${\mathcal R}_{12}={\mathcal R}\otimes 1$, ${\mathcal R}_{23}=1\otimes {\mathcal R}$,
${\mathcal R}_{13}=({\mathfrak p}\otimes 1)\, {\cal R}_{23}$. 
The Cartan part of the universal R-matrix can be isolated as 
${\mathcal R}=
\tilde{\mathcal R}q^{ \frac{ h_{1} \otimes h_{1} }{2}}$, 
where $\tilde{\mathcal R}$ is called the reduced universal R-matrix, which is a series on 
$\{e_{0}\otimes 1 , e_{1} \otimes 1, 1\otimes f_{0}, 1 \otimes f_{1}\}$. 
We symbolically write this as $\tilde{\mathcal R}=
\tilde{\mathcal R}(\{e_{0}\otimes 1 , e_{1} \otimes 1, 1\otimes f_{0}, 1 \otimes f_{1} \})$.  
The Yang-Baxter equation 
\begin{align}
{\mathcal R}_{12}{\mathcal R}_{13}{\cal R}_{23}=
{\mathcal R}_{23}{\mathcal R}_{13}{\mathcal R}_{12} \label{YBE}
\end{align}
follows from the relations \eqref{R-def}. 
We will also use 
\begin{align}
\overline{\mathcal R}={\mathfrak p}{\mathcal R}={\mathcal R}_{21},
\label{uniRb}
\end{align}
which is an element of a completion of  $ {\mathcal B}_{-} \otimes {\mathcal B}_{+} $. 
Taking note on the first relation in \eqref{R-def}, one can show that 
the universal R-matrices commute with the co-multiplication of any Cartan elements
\begin{align}
[{\mathcal R},q^{\xi h_{i}} \otimes q^{\xi h_{i}}]=[\overline{\mathcal R},q^{\xi h_{i}} \otimes q^{\xi h_{i}}]=0 
\quad 
\text{for} \quad \xi \in \mathbb{C}, \quad i=1,2. 
 \label{uniCar}
\end{align}
The following relations follow from the uniqueness of the universal R-matrix \cite{KT92-1} 
and the fact that \eqref{auto1} is an automorphism of each Borel subalgebra.
\begin{align}
(\sigma \otimes \sigma ) {\mathcal R}={\mathcal R}, 
\qquad 
(\sigma \otimes \sigma ) \overline{\mathcal R}=\overline{\mathcal R} .
\label{Rinv}
\end{align} 
\subsection{The quantum algebra $U_{q}(sl_2)$}
The quantum algebra $U_{q}(sl_2)$ is  generated by the elements $E, F, q^{\xi H}$ 
for $\xi \in {\mathbb C}$ obeying the following relations:
\begin{align}
&
q^{0H}=q^{0}=1, \quad 
q^{\xi H}q^{\eta H}=q^{(\xi+\eta )H}, \quad 
q^{\xi H}Eq^{-\xi H}=q^{2\xi}E , \qquad q^{\xi H}Fq^{-\xi H}=q^{-2\xi}F ,
\nonumber
\\ &
[E, F] = \frac{q^{H} - q^{-H} }{q-q^{-1}}, 
\quad \xi,\eta  \in {\mathbb C}.
\label{HEF-sl2}
\end{align}
The upper (resp.\ lower) Borel subalgebra is generated by the elements $E,q^{\xi H}$ (resp.\ $F,q^{\xi H}$). 
The Casimir element 
\begin{align}
C=FE+\frac{q^{H+1} + q^{-H-1}}{(q-q^{-1})^{2}} 
=EF+\frac{q^{H-1} + q^{-H+1}}{(q-q^{-1})^{2}}
\label{Casimir}
\end{align}
is central in $U_{q}(sl_2)$. 
We have an automorphism 
\begin{align}
\sigma(E)= F,
\qquad 
\sigma(F)= E,
\qquad 
\sigma(q^{\xi H})= q^{-\xi H}
\qquad 
\sigma(q)=q,
\end{align}
 and an anti-automorphism 
\begin{align}
& 
E^{t}= q^{-H-1}F,
\qquad 
F^{t}=Eq^{H+1},
\qquad 
(q^{\xi H})^{t}= q^{\xi H}, 
\qquad
q^{t}=q
 \label{t-sl2}
\end{align}
of the algebra.
These are $U_{q}(sl_2)$ analogues  of \eqref{auto1} and \eqref{a-auto1}, respectively. 
There is an algebra homomorphism called evaluation map
\footnote{
We follow \cite{BGKNR10} and consider 
the general gradation of the algebra.} 
$\mathsf{ev}_{x}$:
$U_{q}(\widehat{sl_2}) \mapsto U_{q}(sl_2)$,
\begin{align}
\begin{split}
& e_{0} \mapsto x^{s_{0}}F,  \qquad
 f_{0} \mapsto x^{-s_{0}} E, \qquad
q^{\xi h_{0}}  \mapsto q^{-\xi H},
\\
& e_{1} \mapsto x^{s_{1}} E,  \qquad
f_{1} \mapsto x^{-s_{1}} F, \qquad
q^{\xi h_{1}}  \mapsto q^{\xi H},
\end{split}
\label{eva}
\end{align}
where $x  \in {\mathbb C}^{\times}$ is the spectral parameter. 
We introduce an operation to permute the parameters $s_{0}$ and $s_{1}$:
\begin{align}
\zeta(s_{0})=s_{1}, \quad \zeta(s_{1})=s_{0},
\label{zeta} 
\end{align}
One can verify consistency of these: 
\begin{align}
& \sigma\circ \mathsf{ev}_{x}=(\zeta \cdot \mathsf{ev}_{x}) \circ \sigma  ,
\label{sigev}
 \\
&( \mathsf{ev}_{x}(a))^{t}= \mathsf{ev}_{x^{-1}} (a^{t}) \quad 
 \text{for} \quad  a \in U_{q}(sl_{2}),
 \label{evt}
 \end{align} 
where $\circ$ is composition of maps and $\zeta \cdot \mathsf{ev}_{x} $ is the map \eqref{eva} with the replacement of the parameters $(s_0,s_1) \to (s_1,s_0)$. 
In case the objects  on which \eqref{sigev} is acting do not depend on the parameters $s_0$ and $s_1$, 
\eqref{sigev} reduces to 
$ \sigma\circ \mathsf{ev}_{x}=\zeta \circ \mathsf{ev}_{x} \circ \sigma $. 
The fundamental representation $\pi$ of $U_{q}(sl_2)$ is given by 
$\pi(E)=E_{12}$, $\pi(F)=E_{21}$ and $\pi(q^{\xi H})=q^{\xi}E_{11}+q^{-\xi}E_{22}$ , where
$E_{ij}$ is the $2 \times 2$ matrix unit whose
$(k,l)$-element is $\delta_{i,k}\delta_{j,l}$. 
The composition 
$\pi_{x}=\pi \circ \mathsf{ev}_{x}$
gives the (fundamental) evaluation representation of $U_{q}(\widehat{sl_2})$.
For the fundamental representation,  we define 
 an algebra automorphism $\sigma$ and an algebra anti-automorphism $^{t}$
 of the algebra of $2\times 2$ matrices over ${\mathbb C}$ by 
\begin{align}
\sigma(E_{ij})&=E_{3-i,3-j}, 
\\[6pt]
E_{ij}^{t}&=E_{ji}, \qquad i,j=1,2.
\end{align}
In this case, the anti-homomorphism $^{t}$ coincides with transposition of $2 \times 2$ matrices. 
We have an identity of algebra homomorphisms 
\begin{align}
\pi \circ \sigma = \sigma \circ \pi
\label{pisig}
\end{align}
 and an identity of algebra
 anti-homomorphisms 
 \begin{align}
 \pi (a^{t}) =  (\pi(a))^{t} \quad \text{for} \quad  a \in U_{q}(sl_{2}),
 \label{pit}
 \end{align}
  which justify our use of the same symbol for different maps. 

\subsection{q-oscillator algebras}
We introduce two kinds of oscillator algebras
\footnote{$\mathrm{Osc}_{1}$ is  same as the one  defined by eq.(2.25) in \cite{BT18};
while $\mathrm{Osc}_{2}$ is slightly different from the one defined by eq.(2.26) in \cite{BT18}. 
Let $\hs^{\prime}_{2},\es^{\prime}_{2}, \fs^{\prime}_{2}$ be $\hs_{2},\es_{2}, \fs_{2}$ in eq.(2.26) in \cite{BT18}. 
They are related to $\hs_{2},\es_{2}, \fs_{2}$ in \eqref{osc2} in this paper as
$\hs^{\prime}_{2}=\hs_{2}+2,\es^{\prime}_{2}=q^{\beta}\es_{2}, \fs^{\prime}_{2}=q^{-\beta-2}\fs_{2}$, $\beta \in {\mathbb C}$. 
Moreover, one can define $\mathrm{Osc}_{1}$ as a contraction of $U_{q}(sl_{2})$ (cf.\ \cite{CK90}). 
We define a homomorphism $\rho_{\mu}:  U_{q}(sl_{2}) \to
\mathrm{Osc}_{1} $ (see for example \cite{KS97}) by the relations
\begin{align}
\rho_{\mu}(E)=\es_{1}(q^{\mu}-q^{-\mu -\hs_{1}-2}), 
\quad 
\rho_{\mu}(F)=\fs_{1}, 
\quad 
\rho_{\mu}(q^{\xi H})=q^{\xi (\mu + \hs_{1})}, 
\quad 
\mu, \xi \in \mathbb{C}. 
\label{HPrel}
\end{align}
This realizes the Verma module of $U_{q}(sl_{2})$ with the highest weight $\mu$ on the Fock space. 
The generators of the q-oscillator algebra $\mathrm{Osc}_{1} $ can  be given by contraction of  \eqref{HPrel} 
(see \cite{CK90}, and eq. (2.30) in \cite{BT18}. 
(The similarity transformation by the factor $q^{-\frac{\mu s_{0}\hs_{1}}{2s}} $  
 is related to the change of basis mentioned in footnote 7 in \cite{BT18}.))
\begin{multline}
\es_{1}= \lim_{q^{-\mu} \to 0 }
q^{-\frac{\mu s_{0}\hs_{1}}{2s}} \rho_{\mu}(E)q^{\frac{\mu s_{0}\hs_{1}}{2s}} q^{-\frac{\mu s_{1}}{s}}, \qquad 
\fs_{1}= \lim_{q^{-\mu} \to 0 }
 q^{-\frac{\mu s_{0}\hs_{1}}{2s}} \rho_{\mu}(F) q^{\frac{\mu s_{0}\hs_{1}}{2s}} q^{-\frac{\mu s_{0}}{s}},
\\
q^{\xi \hs_{1}}= \lim_{q^{-\mu} \to 0 } q^{-\frac{\mu s_{0}\hs_{1}}{2s}} \rho_{\mu}(q^{\xi (H-\mu )}) q^{\frac{\mu s_{0}\hs_{1}}{2s}}, 
\label{limosc1}
\end{multline}
where the limit is taken with respect to $\mu$ ($q$ is constant). 
}
 $\mathrm{Osc}_{i}$ ($i=1,2$).
 They are generated by  the elements $\es_{i}, \fs_{i},q^{\xi \hs_{i}}$ (for $\xi \in {\mathbb C}$) 
 obeying the following relations:
\begin{align}
\begin{split}
& q^{0\hs_{1}}=q^{0}=1, \quad q^{\xi \hs_{1}} q^{\eta \hs_{1}}=q^{(\xi +\eta) \hs_{1}}
\quad q^{\xi \hs_{1}}\es_{1} q^{-\xi \hs_{1}} =q^{2\xi }\es_{1},
\quad  q^{\xi \hs_{1}}\fs_{1} q^{-\xi \hs_{1}} =q^{-2\xi }\fs_{1},
\\
& \fs_{1}\es_{1}=q\frac{1-q^{\hs_{1}}}{( q-q^{-1})^{2} } ,
\qquad
\es_{1}\fs_{1}=q\frac{1-q^{\hs_{1}-2 }}{( q-q^{-1})^{2} },
\quad \xi, \eta \in {\mathbb C},
\end{split}
\label{osc1}
\end{align}
\begin{align}
\begin{split}
& q^{0\hs_{2}}=q^{0}=1, \quad q^{\xi \hs_{2}} q^{\eta \hs_{2}}=q^{(\xi +\eta) \hs_{2}}
\quad q^{\xi \hs_{2}}\es_{2} q^{-\xi \hs_{2}} =q^{2\xi }\es_{2},
\quad  q^{\xi \hs_{2}}\fs_{2} q^{-\xi \hs_{2}} =q^{-2\xi }\fs_{2},
\\
& \fs_{2}\es_{2}=q\frac{1-q^{-\hs_{2}-2}}{( q-q^{-1})^{2} } ,
\qquad
\es_{2}\fs_{2}=q\frac{1-q^{-\hs_{2} }}{( q-q^{-1})^{2} }, 
\quad \xi, \eta \in {\mathbb C}.
\end{split}
\label{osc2}
\end{align}
Note that $\mathrm{Osc}_{2}$ can be realized in terms of $\mathrm{Osc}_{1}$:
\begin{align}
\es_{2}=\fs_{1}, 
\qquad 
\fs_{2}=\es_{1},
\qquad 
q^{\xi \hs_{2}}=q^{- \xi \hs_{1}}.
\label{osc12}
\end{align}
The following relations follow from  \eqref{osc1} and \eqref{osc2}:
\begin{align}
& [\es_{1},\fs_{1}]=\frac{q^{\hs_{1} }}{q-q^{-1}},
\qquad
[\es_{2},\fs_{2}]=-\frac{q^{-\hs_{2} }}{q-q^{-1}},
\label{comosc1}
\\
& [\es_{1},\fs_{1}]_{q^{-2}}= \frac{1}{q-q^{-1}},
\qquad
[\es_{2},\fs_{2}]_{q^{2}} = -\frac{q^{2}}{q-q^{-1}}.
\label{comosc2}
\end{align}
We will use anti-involutions of $\mathrm{Osc}_{i}$ 
(analogues of \eqref{t-sl2}) 
defined by
\begin{align}
& 
\es^{t}_{i} = q^{-\hs_{i}-1}\fs_{i},
\qquad 
\fs^{t}_{i} = \es_{i}q^{\hs_{i}+1},
\qquad 
(q^{\xi \hs_{i}})^{t} = q^{\xi \hs_{i}}, 
 \label{t-osc}
\end{align}
where $(ab)^{t}=b^{t}a^{t}$ holds for any $a,b \in \mathrm{Osc}_{i}$, $i=1,2$.
We define 
the homomorphism $\rho_x^{(i)}: \mathcal{B}_{+} \to
\mathrm{Osc}_{i}, i=1,2 $ by the relations
\begin{align}\label{rho+}
& \rho^{(i)}_{x}(e_{0})=x^{s_{0}}\fs_{i}, 
\quad \rho^{(i)}_{x}(e_{1})=x^{s_{1}}\es_{i}, 
\quad \rho^{(i)}_{x}(q^{\xi h_{0}})=q^{-\xi \hs_{i}}, \quad \rho^{(i)}_{x}(q^{\xi h_{1}})=q^{\xi \hs_{i}}, 
\end{align}
or  the homomorphism $\rho_x^{(i)}: \mathcal{B}_{-} \to
\mathrm{Osc}_{i}, i=1,2 $  by the relations
\begin{align}\label{rho-}
&\rho^{(i)}_{x}(f_{0})=x^{-s_{0}}\es_{i}, 
\quad \rho^{(i)}_{x}(f_{1})=x^{-s_{1}}\fs_{i}, 
\quad \rho^{(i)}_{x}(q^{\xi h_{0}})=q^{-\xi \hs_{i}}, \quad \rho^{(i)}_{x}(q^{\xi h_{1}})=q^{\xi \hs_{i}}. 
\end{align}
These maps are related each other as (cf.\ \eqref{osc12})
\begin{align}
\rho^{(2)}_{x}=(\zeta \cdot \rho^{(1)}_{x}) \circ \sigma .
 \label{rh12}
\end{align}
Here $\zeta \cdot \rho^{(1)}_{x}$ is the map \eqref{rho+} or \eqref{rho-} with the replacement of the parameters $(s_0,s_1) \to (s_1,s_0)$. 
In case the objects  on which \eqref{rh12} is acting do not depend on the parameters $s_0$ and $s_1$, 
\eqref{rh12} reduces to 
$ \rho^{(2)}_{x}=\zeta \circ \rho^{(1)}_{x}\circ \sigma $. 

\section{L-operators}
In this section, we review various L-operators, which are building blocks of T-and Q-operators. 
They are holomorphic images of the universal R-matrix in various representations of 
 Borel subalgebras of $U_{q}(\widehat{sl_2})$. We will make use of 
  the product expression of the universal R-matrix given by 
Khoroshkin and Tolstoy \cite{TK92,KT94}, which is reviewed in Appendix A. 
Their universal R-matrix was already reviewed by several authors 
(see for example, \cite{ZG93,KST94,BGKNR10,BGKNR12,MT15,BT18,PSZ16}). 
In particular,  a pedagogical account on how to evaluate it in the context of 
Baxter Q-operators can be found in \cite{BGKNR10,BGKNR12}. 
\subsection{L-operators for T-operators}
We define the universal L-operators  by 
\begin{align}
{\mathcal L} (x)&= 
(\pi_{x} \otimes 1 ) \mathcal{R}, 
\qquad 
\overline{\mathcal L} (x)= 
(\pi_{x} \otimes 1 ) \overline{\mathcal R}, 
\quad 
x \in {\mathbb C} .
\end{align}
Evaluating the first components of the universal R-matrices in the fundamental evaluation representation $\pi_{x}$, 
we obtain
\begin{multline}
{\mathcal L} (x)= \left(1+ \lambda E_{12}\otimes \sum_{k=0}^{\infty}(-q^{-1})^{k}x^{ks+s_{1}}f_{\alpha+k\delta}\right)
\\
\times  \left(E_{11}\otimes \exp\left(- \lambda \sum_{k=1}^{\infty}\frac{(-x^{s})^{k} }{q^{k}+q^{-k}}f_{k\delta} \right) +
E_{22}\otimes \exp\left( \lambda \sum_{k=1}^{\infty}\frac{(-q^{-2}x^{s})^{k} }{q^{k}+q^{-k}}f_{k\delta} \right)  \right)
\\
\times  \left(1+ \lambda E_{21}\otimes \sum_{k=0}^{\infty}(-q^{-1})^{k}x^{ks+s_{0}}f_{\delta-\alpha+k\delta}\right)
\left(E_{11} \otimes q^{\frac{h_{1}}{2}}+E_{22} \otimes q^{-\frac{h_{1}}{2}}\right),
\end{multline}
\begin{multline}
\overline{\mathcal L} (x)= \left(1+ \lambda E_{21}\otimes \sum_{k=0}^{\infty}(-q)^{k}x^{-ks-s_{1}}e_{\alpha+k\delta}\right)
\\
\times  \left(E_{11}\otimes \exp\left(- \lambda \sum_{k=1}^{\infty}\frac{(-x^{-s})^{k} }{q^{k}+q^{-k}}e_{k\delta} \right) +
E_{22}\otimes \exp\left( \lambda \sum_{k=1}^{\infty}\frac{(-q^{2}x^{-s})^{k} }{q^{k}+q^{-k}}e_{k\delta} \right)  \right)
\\
\times  \left(1+ \lambda E_{12}\otimes \sum_{k=0}^{\infty}(-q)^{k}x^{-ks-s_{0}}e_{\delta-\alpha+k\delta}\right)
\left(E_{11} \otimes q^{\frac{h_{1}}{2}}+E_{22} \otimes q^{-\frac{h_{1}}{2}}\right).
\end{multline}
Here the root vectors 
$\{ e_{\alpha+k\delta},e_{k\delta},e_{\delta-\alpha+k\delta}, 
f_{\alpha+k\delta},f_{k\delta},f_{\delta-\alpha+k\delta} \}$ 
 can be expressed in terms of the basic generators $e_{\alpha } =e_{1}$, $e_{\delta- \alpha } =e_{0}$, 
$f_{\alpha } =f_{1}$, $f_{\delta- \alpha } =f_{0}$ via \eqref{root1} and \eqref{root2}.  
Evaluating the second components of these universal L-operators in the fundamental evaluation representation 
$\pi_{1}=\pi_{x}|_{x=1}$, 
we obtain the R-matrices of the 6-vertex model.
\begin{align}
R(x)=
q^{\frac{1}{2}}\phi(x)(1 \otimes \pi_{1}){\mathcal L} (x)& =
\begin{pmatrix}
q-q^{-1}x^{s} & 0 & 0 & 0 \\
0 & 1-x^{s} &  \lambda x^{s_{1}} & 0 \\
0 & \lambda x^{s_{0}} & 1-x^{s}  & 0 \\
 0 & 0 & 0 & q-q^{-1}x^{s}
 \end{pmatrix} ,
 \label{Rmat1}
 \\[6pt]
\overline{R}(x)=
q^{\frac{1}{2}}\phi(x^{-1})(1 \otimes \pi_{1})\overline{\mathcal L} (x)& =
\begin{pmatrix}
q-q^{-1}x^{-s} & 0 & 0 & 0 \\
0 & 1-x^{-s} &  \lambda x^{-s_{0}} & 0 \\
0 & \lambda x^{-s_{1}} & 1-x^{-s}  & 0 \\
 0 & 0 & 0 & q-q^{-1}x^{-s}
 \end{pmatrix} ,
 \label{Rmat2}
\end{align}
where the overall factor is defined by $\phi(x)=e^{-\Lambda(x^{s}q^{-1})} $, 
$\Lambda (x)=\sum_{k=1}^{\infty}\frac{q^{2k}+q^{-2k}}{k(q^{k}+q^{-k})}x^{k}$. 
\subsection{L-operators for Q-operators}
We define the universal L-operators for Q-operators  by
\begin{align}
{\mathcal L}^{(a)} (x)&= 
(\mathsf{\rho}^{(a)}_{x} \otimes 1 ) \mathcal{R}, 
\qquad
\overline{\mathcal L}^{(a)} (x)= 
(\mathsf{\rho}^{(a)}_{x} \otimes 1 ) \overline{\mathcal R}, 
\quad
x \in {\mathbb C}, \quad a=1,2.
\end{align}
One can calculate these based on the explicit expression of the universal R-matrix in Appendix A. 
In particular, $\mathcal{L}^{(1)}(x)$ and $\overline{\mathcal L}^{(2)}(x)$ have simple expressions:
\begin{align}
&\mathcal{L}^{(1)}(x)= \exp_{q^{-2}} 
\left( \lambda x^{s_{1}} \es_{1} \otimes f_{\alpha } \right) 
\exp  
\left(  \sum_{k=1}^{\infty} \frac{(-1)^{k-1} x^{sk}}{[2k]_{q}} \otimes f_{ k \delta } \right) 
 \exp_{q^{-2}} 
\left( \lambda x^{s_{0}} \fs_{1} \otimes f_{\delta-\alpha  } \right) 
 q^{\frac{1}{2}\hs_{1}\otimes h_{1}} ,
 \label{UL1}
\\
&\overline{\mathcal L}^{(2)}(x)= 
\nonumber \\
&=\exp_{q^{-2}} 
\left( \lambda x^{-s_{1}} \fs_{2} \otimes e_{\alpha } \right) 
\exp  
\left(  \sum_{k=1}^{\infty} \frac{(-q^{2}x^{-s})^{k} }{[2k]_{q}} \otimes e_{ k \delta } \right) 
 \exp_{q^{-2}} 
\left( \lambda x^{-s_{0}} \es_{2} \otimes e_{\delta-\alpha  } \right) 
q^{\frac{1}{2}\hs_{2}\otimes h_{1}} .
 \label{ULb2}
 \end{align}
Direct evaluations of  $\mathcal{L}^{(2)}(x)$ and $\overline{\mathcal L}^{(1)}(x)$ contain infinite products
\footnote{See eq. (B.21) in \cite{BT18}.}
of q-exponential functions (at least in the root ordering which we have adapted). 
In order to avoid complicated expressions, we use 
the relations \eqref{Rinv}, \eqref{rh12}, \eqref{osc12} and $\sigma=\sigma^{-1}$ for \eqref{UL1} and \eqref{ULb2}, to get 
\begin{align}
&\mathcal{L}^{(2)}(x)= \zeta \circ (1 \otimes \sigma ) \mathcal{L}^{(1)}(x)=
\nonumber \\
&=
\exp_{q^{-2}} 
\left( \lambda x^{s_{0}} \fs_{2} \otimes f_{\delta-\alpha } \right) 
\exp  
\left(  \sum_{k=1}^{\infty} \frac{(-q^{2} x^{s})^{k}}{[2k]_{q}} \otimes \overline{f}_{ k \delta } \right) 
 \exp_{q^{-2}} 
\left( \lambda x^{s_{1}} \es_{2} \otimes f_{\alpha  } \right) 
q^{\frac{1}{2}\hs_{2}\otimes h_{1}} ,
 \label{UL2}
\\
&\overline{\mathcal L}^{(1)}(x)= \zeta \circ (1 \otimes \sigma ) \overline{\mathcal L}^{(2)}(x)=
\nonumber \\
&= \exp_{q^{-2}} 
\left( \lambda x^{-s_{0}} \es_{1} \otimes e_{\delta-\alpha } \right) 
\exp  
\left(  \sum_{k=1}^{\infty} \frac{(-1)^{k-1} x^{-sk} }{[2k]_{q}} \otimes \overline{e}_{ k \delta } \right) 
 \exp_{q^{-2}} 
\left( \lambda x^{-s_{1}} \fs_{1} \otimes e_{\alpha  } \right) 
q^{\frac{1}{2}\hs_{1}\otimes h_{1}} ,
 \label{ULb1}
 \end{align}
 where the root vectors $ \overline{e}_{ k \delta }$ and $ \overline{f}_{ k \delta }$, 
 which can be expressed in terms of the basic generators $e_{1}$, $e_{0}$, $f_{1}$, $f_{0}$, 
 come from \eqref{sigroots}. 
 Now we evaluate the second component of the universal L-operators in the 
 fundamental evaluation representation $\pi_{1}=\pi_{x}|_{x=1}$. 
 We  normalize the L-operators as
\begin{align}
\Lf^{(i)}(x)&= \phi^{(1)}(x)( 1 \otimes \pi_{1} ) {\mathcal L}^{(i)}(x) ,  \quad 
\Lfb^{(i)}(x)= \phi^{(1)}(x^{-1})( 1 \otimes \pi_{1} ) \overline{\mathcal L}^{(i)}(x) , 
\end{align}
where
 $ \phi^{(1)}(x)=e^{-\Phi (x^{s})} $, 
$\Phi (x)=\sum_{k=1}^{\infty}\frac{1}{k(q^{k}+q^{-k})}x^{k}$, $i=1,2$.
These are L-operators for Q-operators for the $XXZ$-spin chain. 
Explicitly, one obtains 
\footnote{The L-operators with the superscript `$^{(1)}$' are in the same convention as the ones in \cite{BT18}, 
but the ones with `$^{(2)}$' are superficially 
 different since we have slightly changed the definition of 
the q-oscillator algebra $\mathrm{Osc}_{2}$.}
\begin{align}
\Lf^{(1)}(x)& 
=
\begin{pmatrix}
q^{\frac{\hs_{1}}{2}}  & \lambda x^{s_{0}} \fs_{1} q^{-\frac{\hs_{1}}{2} } \\
\lambda x^{s_{1}} \es_{1} q^{\frac{\hs_{1}}{2} } & q^{- \frac{\hs_{1}}{2}} -q^{-1} x^{s} q^{\frac{\hs_{1}}{2}}
 \end{pmatrix} ,
 \label{LQ1}
\\[6pt]
\Lf^{(2)}(x)& = 
\begin{pmatrix}
q^{\frac{\hs_{2}}{2}} -q^{-1} x^{s} q^{-\frac{\hs_{2}}{2}}
 & \lambda x^{s_{0}} \fs_{2} q^{-\frac{\hs_{2}}{2} } \\
\lambda x^{s_{1}} \es_{2} q^{\frac{\hs_{2}}{2} } & q^{- \frac{\hs_{2}}{2}} 
 \end{pmatrix} ,
 \label{LQ2}
\\
\overline{\Lf}^{(1)}(x)
&=
\begin{pmatrix}
q^{\frac{\hs_{1}}{2}}  & \lambda x^{-s_{1}} \fs_{1} q^{-\frac{\hs_{1}}{2} } \\
\lambda x^{-s_{0}} \es_{1} q^{\frac{\hs_{1}}{2} } & q^{- \frac{\hs_{1}}{2}} -q^{-1} x^{-s} q^{\frac{\hs_{1}}{2}}
 \end{pmatrix} ,
 \label{LhQ1}
\\[6pt]
\overline{\Lf}^{(2)}(x)
&=
\begin{pmatrix}
q^{\frac{\hs_{2}}{2}} -q^{-1} x^{-s} q^{-\frac{\hs_{2}}{2}}
 & \lambda x^{-s_{1}} \fs_{2} q^{-\frac{\hs_{2}}{2} } \\
\lambda x^{-s_{0}} \es_{2} q^{\frac{\hs_{2}}{2} } & q^{- \frac{\hs_{2}}{2}} 
 \end{pmatrix} .
 \label{LhQ2}
\end{align}
In addition to the above L-operators, we need L-operators proportional to the inverse of them. 
\footnote{
We could interpret these as follows (cf. \cite{BT18}). Consider universal L-operators of the form: 
\begin{align}
\check{\mathcal{L}}^{(a)}(x)=(\rho_x^{(a)} \otimes 1)
 \overline{\mathcal R}^{-1}=
 \overline{\mathcal{L}}^{(a)}(x)^{-1}, 
\quad 
\check{\overline{\mathcal{L}}}^{(a)}(x)=(\rho_x^{(a)} \otimes 1) {\mathcal R}^{-1}
={\mathcal L}^{(a)}(x)^{-1}, \quad a=1,2.
 \label{uni-Lop2}
\end{align}
The L-operators \eqref{LQ1-an}-\eqref{LQ2-an} are images of these:
\begin{align}
\Lfc^{(i)}(x)&= \Check{\phi}^{(1)}(x^{-1})( 1 \otimes \pi_{1} ) \Check{\mathcal L}^{(i)}(x) , \quad 
\Lfcb^{(i)}(x)= \Check{\phi}^{(1)}(x)( 1 \otimes \pi_{1} )
\Check{\overline{\mathcal L}}^{(i)}(x) 
,
\end{align} 
where 
$ \Check{\phi}^{(1)}(x)=(-x^{-s}q^{-1})e^{-\Phi (x^{s}q^{2})} $. 
There is a useful identity
$  \phi^{(1)}(x) \Check{\phi}^{(1)}(x)=1-q^{-1}x^{-s}$.
}
\begin{align}
\Lfc^{(1)}(x)
&=
\begin{pmatrix}
q^{\frac{\hs_{1}}{2}}   -q^{-1} x^{s} q^{-\frac{\hs_{1}}{2}}& \lambda x^{s_{0}} \fs_{1} q^{-\frac{\hs_{1}}{2} } \\
\lambda x^{s_{1}} \es_{1} q^{\frac{\hs_{1}}{2} } &  -q^{-1} x^{s} q^{\frac{\hs_{1}}{2}}
 \end{pmatrix} ,
 \label{LQ1-an}
\\[6pt]
\Lfcb^{(1)}(x)& =
\begin{pmatrix}
q^{\frac{\hs_{1}}{2}}   -q^{-1} x^{-s} q^{-\frac{\hs_{1}}{2}} 
& \lambda x^{-s_{1}} \fs_{1} q^{-\frac{\hs_{1}}{2} } \\
\lambda x^{-s_{0}} \es_{1} q^{\frac{\hs_{1}}{2} } & 
 -q^{-1} x^{-s} q^{\frac{\hs_{1}}{2}}
 \end{pmatrix} ,
 \label{LhQ1-an}
 \\[6pt]
\Lfc^{(2)}(x)
&=
\begin{pmatrix}
   -q^{-1} x^{s} q^{-\frac{\hs_{2}}{2}}& \lambda x^{s_{0}} \fs_{2} q^{-\frac{\hs_{2}}{2} } \\
\lambda x^{s_{1}} \es_{2} q^{\frac{\hs_{2}}{2} } & q^{-\frac{\hs_{2}}{2}} -q^{-1} x^{s} q^{\frac{\hs_{2}}{2}}
 \end{pmatrix} ,
 \label{LQ2-an}
\\[6pt] 
\Lfcb^{(2)}(x)& =
\begin{pmatrix}
  -q^{-1} x^{-s} q^{-\frac{\hs_{2}}{2}} 
& \lambda x^{-s_{1}} \fs_{2} q^{-\frac{\hs_{2}}{2} } \\
\lambda x^{-s_{0}} \es_{2} q^{\frac{\hs_{2}}{2} } & 
q^{-\frac{\hs_{2}}{2}}  -q^{-1} x^{-s} q^{\frac{\hs_{2}}{2}}
 \end{pmatrix} .
 \label{LhQ2-an}
 \end{align}
The L-operators with the superscript `$^{(2)}$' can be obtained from the ones with `$^{(1)}$'. 
\begin{align}
\Lf^{(2)}(x)&=\zeta \circ (1 \otimes  \sigma ) \Lf^{(1)}(x), 
\label{L2L1} \\
\overline{\Lf}^{(2)}(x)&=\zeta \circ (1 \otimes  \sigma ) \overline{\Lf}^{(1)}(x), 
\label{L2L1b} \\
\Lfc^{(2)}(x)&=\zeta \circ (1 \otimes  \sigma )\Lfc^{(1)}(x), 
\label{L2L1c}
\\
\Lfcb^{(2)}(x)&=\zeta \circ (1 \otimes  \sigma )\Lfcb^{(1)}(x).
\label{L2L1cb}
\end{align}
 One can check that 
these L-operators satisfy
\begin{align}
\Lf^{(a)}(x)\Lfcb^{(a)}(x)&=\Lfcb^{(a)}(x)\Lf^{(a)}(x)=1-q^{-1}x^{-s},
\label{LLcb1=c}
 \\
 \Lfc^{(a)}(x)\Lfb^{(a)}(x)&=\Lfb^{(a)}(x)\Lfc^{(a)}(x)=1-q^{-1}x^{s},
 \label{LcLb1=c}
 \\
g_{2}\Lf^{(a)} (xq^{\frac{4}{s}})^{t_{2}} g_{2}^{-1}\Lfcb^{(a)} (x)^{t_{2}}&=
\Lfcb^{(a)} (x)^{t_{2}} g_{2}\Lf^{(a)} (xq^{\frac{4}{s}})^{t_{2}} g_{2}^{-1}=
q^{2}-q^{-1}x^{-s} ,
 \label{LcbL1s=c}
 \\
g_{2}\Lfc^{(a)} (xq^{\frac{4}{s}})^{t_{2}} g_{2}^{-1}\Lfb^{(a)} (x)^{t_{2}}&=
\Lfb^{(a)} (x)^{t_{2}} g_{2}\Lfc^{(a)} (xq^{\frac{4}{s}})^{t_{2}} g_{2}^{-1}=
q^{2}-q^{3}x^{s} ,
\label{LbLc1s=c}
\end{align}
where $a=1,2$, 
$g=\mathrm{diag}(q^{\frac{s_{0}-s_{1}}{s}},q^{-\frac{s_{0}-s_{1}}{s}})$, 
$g_{2}=1 \otimes g$, and $^{t_{2}}$
 is the transposition in the second component of the tensor product.
Note that the matrix $g$ is invariant under the map $\zeta \circ \sigma$: 
\begin{align}
\zeta \circ \sigma (g)=g .
\label{ginv}
\end{align}
One can also check the following relations for the L-operators: 
\begin{align}
\Lf^{(a)}(x)^{t_{1}t_{2}}&=\Lfb^{(a)}(x^{-1}), &
\Lfb^{(a)}(x)^{t_{1}t_{2}}&=\Lf^{(a)}(x^{-1}), 
\nonumber \\
\Lfc^{(a)}(x)^{t_{1}t_{2}}&=\Lfcb^{(a)}(x^{-1}), &
\Lfcb^{(a)}(x)^{t_{1}t_{2}}&=\Lfc^{(a)}(x^{-1}),
\quad a=1,2,
\label{tt}
\end{align}
where  $^{t_{1}}$
 is the anti-involution \eqref{t-osc} in the first component of the tensor product.
The relations 
\eqref{LLcb1=c} and \eqref{LcLb1=c}, and 
\eqref{LcbL1s=c} and \eqref{LbLc1s=c} can be interchanged by \eqref{tt}. 
Moreover \eqref{LLcb1=c}-\eqref{LbLc1s=c} and \eqref{tt} for $a=2$ follow from the ones for $a=1$ via 
\eqref{L2L1}-\eqref{L2L1cb} and \eqref{ginv}. 
%
\section{Reflection equation and K-operators}
A systematic approach for construction of quantum integrable systems with open 
boundaries was developed by Sklyanin \cite{Skly}. The key equation 
for this is the reflection equation (boundary Yang-Baxter equation) \cite{Ch84}.
We start from the following form of the reflection equation 
and the dual reflection equation for the R-matrices 
\eqref{Rmat1} and \eqref{Rmat2}:
\begin{align}
R_{12} \left(\frac{y}{x}\right) K_{1}(x) \overline{R}_{12} \left( xy \right) 
 K_{2}(y) 
&=K_{2}(y) 
 R_{12} \left(\frac{1}{xy} \right)  K_{1}(x) \overline{R}_{12} \left(\frac{x}{y}\right),
\label{refeq0}
\\
R_{12}\left(\frac{y}{x}\right) \overline{K}_{1}(x)^{t_{1}} 
g_{2} \overline{R}_{12}\left( xyq^{-\frac{4}{s}} \right) g_{2}^{-1}
\overline{K}_{2}(y)^{t_{2}} 
&=\overline{K}_{2}(y)^{t_{2}}
g_{2}^{-1} R_{12}\left(\frac{q^{\frac{4}{s}}}{xy}\right)  g_{2}
\overline{K}_{1}(x)^{t_{1}} \overline{R}_{12}\left(\frac{x}{y}\right),
\label{refeqdual0-2}
\end{align}
where $x,y \in {\mathbb C}^{\times}$, $K_{1}(x)=K(x) \otimes 1$, $K_{2}(y)=1 \otimes K(y)$. 
The most general non-diagonal 
$2\times 2$ matrix solutions of the reflection equations are known in \cite{DeV,GZ,MN}.  
The diagonal solutions  of \eqref{refeq0}  and  \eqref{refeqdual0-2}  
are  specialization of  them:  
\begin{align}
K (x) & =
\begin{pmatrix}
x^{s_{0}} \epsilon_{+}  +  x^{-s_{1}} \epsilon_{-} & 0  \\
0  & x^{-s_{0}}\epsilon_{+}  +  x^{s_{1}} \epsilon_{-}
 \end{pmatrix} ,
 \label{Kmat}
\\
\overline{K}(x)& =
\begin{pmatrix}
q^{-1}x^{s_{0}}\overline{\epsilon}_{+} + qx^{-s_{1}}\overline{\epsilon}_{-} & 
0 \\
0 & 
qx^{-s_{0}}\overline{\epsilon}_{+} + q^{-1}x^{s_{1}}\overline{\epsilon}_{-}
 \end{pmatrix} ,
 \label{Kmat-d}
 \end{align}
where  $\epsilon_{\pm}$ and $\overline{\epsilon}_{\pm}$ are scalar parameters
\footnote{Up to an overall factor, $\overline{K}(xq^{\frac{4}{s}})g^{-2}$ coincides with $\overline{K}(x)$ in eq.\ (4.30) in \cite{BT18}.}. 
We assume  $\epsilon_{+}\epsilon_{-}\overline{\epsilon}_{+}\overline{\epsilon}_{-} \ne 0$ since we
 will deal with solutions which contain $\epsilon_{+}^{-1}, \epsilon_{-}^{-1}, \overline{\epsilon}_{+}^{-1}$ or $\overline{\epsilon}_{-}^{-1}$. 
We remark that these solutions \eqref{Kmat} and \eqref{Kmat-d} are related to each other by the following transformation \cite{MN}:
\begin{align}
\overline{K}(x) &= K^t\left(xq^{-\frac{2}{s}}\right)g|_{\epsilon_{\pm}=\overline{\epsilon}_{\pm}}. 
\end{align}
In addition to \eqref{zeta}, we assume
\begin{align}
\zeta(\epsilon_{+})=\epsilon_{-}, \quad \zeta(\epsilon_{-})=\epsilon_{+}, \quad 
\zeta(\overline{\epsilon}_{+})=\overline{\epsilon}_{-}, \quad 
\zeta(\overline{\epsilon}_{-})=\overline{\epsilon}_{+}.
\label{zeta2}
\end{align}
The K-matrices \eqref{Kmat} and \eqref{Kmat-d} are invariant under the operation $\zeta \circ \sigma$:
\begin{align}
\zeta \circ \sigma (K(x))=K(x), 
\qquad 
\zeta \circ \sigma (\overline{K}(x))=\overline{K}(x).
\label{inv1}
\end{align}
Next, we consider the 
reflection equations and the dual reflection equations for the L-operators 
for Q-operators \eqref{LQ1}, \eqref{LhQ1}, \eqref{LQ1-an} and \eqref{LhQ1-an}: 
\begin{align}
& \Lf^{(a)}_{12} \left(\frac{y}{x}\right) \Kf^{(a)}_{1}(x) \overline{\Lf}^{(a)}_{12} \left( xy \right) 
 K_{2}(y) 
 =K_{2}(y) 
 \Lf^{(a)}_{12} \left(\frac{1}{xy} \right)  \Kf^{(a)}_{1}(x) \overline{\Lf}^{(a)}_{12} \left(\frac{x}{y}\right),
\label{refeqlim1st}
\\
& \Lfc^{(a)}_{12} \left(\frac{y}{x}\right) \Kfcb^{(a)}_{1}(x)^{t_{1}} 
g_{2} \Lfcb^{(a)}_{12}\left( xy q^{-\frac{4}{s}}\right) g_{2}^{-1}
\overline{K}_{2}(y)^{t_{2}} 
=
\nonumber \\
& \hspace{100pt} =\overline{K}_{2}(y)^{t_{2}}
g_{2}^{-1} \Lfc^{(a)}_{12} \left(\frac{q^{\frac{4}{s}}}{xy}\right)  g_{2}
\Kfcb^{(a)}_{1}(x)^{t_{1}} \Lfcb^{(a)}_{12} \left(\frac{x}{y}\right), 
\quad a=1,2.
\label{refeqlim2ndch-pr}
\end{align}
We have solutions (K-operators) \cite{BT18}
\footnote{$\check{\overline{\kappa}}^{(1)}(x) \Kfcb^{(1)}(x)$ corresponds to $\Kfcb^{(1)\prime}(x)$ in page 434 in \cite{BT18}.}
 of these equations for $a=1$.
\begin{align}
\Kf^{(1)}(x)&=\kappa^{(1)}(x)^{-1} x^{s_{0}\hs_{1}} \exp^{-1}_{q^{-2}}\left( -\frac{\epsilon_{-}x^{s}q^{-\hs_{1}}}{\lambda \epsilon_{+}} \right),
\label{Kop1}
\\
\Kfcb^{(1)}(x)&=\check{\overline{\kappa}}^{(1)}(x)^{-1} x^{s_{0}\hs_{1}}q^{-\hs_{1}} \exp_{q^{-2}}
\left( -\frac{\overline{\epsilon}_{-}x^{-s}q^{2-\hs_{1}}}{\lambda \overline{\epsilon}_{+}} \right),
\label{Kop2}
\end{align}
where the normalization functions are defined  by 
\begin{align}
\kappa^{(1)}(x)&=\exp^{-1}_{q^{-2}}\left( -\frac{\epsilon_{-}x^{s}}{\lambda \epsilon_{+}} \right),
\\
\check{\overline{\kappa}}^{(1)}(x)&=\exp_{q^{-2}}\left( -\frac{\overline{\epsilon}_{-}x^{-s}}{\lambda \overline{\epsilon}_{+}} \right).
\end{align}
We remark that the same type of normalization is used in \cite{VW20}. 
The solutions \eqref{Kop1} and \eqref{Kop2} are related to each other by the following transformation:
\begin{align}
\Kfcb^{(1)}(x) &=\left(1+ \frac{\overline{\epsilon}_{-}x^{-s}q}{ \overline{\epsilon}_{+}} \right)^{-1} \Kf^{(1)t}(x^{-1}q^{\frac{2}{s}})^{-1} q^{\frac{s_{0}-s_{1}}{s}\hs_{1}}|_{\epsilon_{\pm}=\overline{\epsilon}_{\pm}}. 
\end{align}
Solutions for $a=2$ follow from the first ones \eqref{Kop1} and \eqref{Kop2}:
\begin{align}
\Kf^{(2)}(x)&=\zeta(\Kf^{(1)}(x)),
\qquad 
\Kfcb^{(2)}(x)=\zeta(\Kfcb^{(1)}(x)).
\label{Kop3}
\end{align}
One can check this by applying $\zeta \circ (1 \otimes \sigma)$ to \eqref{refeqlim1st} and \eqref{refeqlim2ndch-pr} 
for $a=1$ 
(with the help of \eqref{L2L1}-\eqref{L2L1cb}, \eqref{ginv} and \eqref{inv1}).
%
\section{Universal Baxter TQ-relation}
In this section, we apply a universal version of Sklyanin's dressing method \cite{Skly} 
to the K-operators in the previous section and 
obtain more general solutions of the reflection equation. Then we define universal T-and Q-operators for open boundaries integrable systems \cite{BT18}, 
which are elements in $U_{q}(sl_2)$. We will present the universal TQ-relations among them. 

We  define the universal dressed K-operator for a T-operator by 
\begin{align}
{\mathcal K}(x)&={\mathcal L}(x^{-1}) (K(x)\otimes 1) \overline{\mathcal L}(x) , 
\qquad x \in \mathbb{C}^{\times}.
 \label{dressed-UKti}
\end{align}
One can show that \eqref{dressed-UKti} satisfies the following universal  dressed reflection equation 
for a T-operator
 (see Appendix C).
\begin{align}
R_{12} \left(x^{-1} y \right) {\mathcal K}_{13}(x) \overline{R}_{12} \left( xy \right) 
{\mathcal K}_{23}(y) 
&={\mathcal K}_{23}(y)
 R_{12} \left(x^{-1}y^{-1} \right) {\mathcal K}_{13}(x) \overline{R}_{12} \left(x y^{-1} \right), 
 \quad x,y \in \mathbb{C}^{\times}.
\label{RE-dressLti}
\end{align}
We define the universal T-operator  by 
\begin{align}
{\mathcal T}(x) & = 
(\mathrm{tr} \otimes 1)
\left(
(\overline{K}(x^{-1})  \otimes 1)
{\mathcal K}(x)
\right) ,
\qquad x \in \mathbb{C}^{\times},
\label{T-op-fin}
\end{align}
where the trace is taken over the space $\mathrm{End}(\mathbb{C}^{2})$. 
This is invariant under the operation $\zeta \circ \sigma$:
\begin{align}
\zeta \circ \sigma({\mathcal T}(x))={\mathcal T}(x) .
\label{invT}
\end{align}
One can show this by using the relations:  
$(1 \otimes \sigma ) \Lc(x)=\zeta \circ (\sigma \otimes 1)\Lc(x)  $, 
$(1 \otimes \sigma ) \Lcb(x)=\zeta \circ (\sigma \otimes 1)\Lcb(x)  $ 
(these follow from \eqref{Rinv}, \eqref{sigev}, \eqref{pisig}), \eqref{inv1} 
and $\mathrm{tr}\, \sigma(A)=\mathrm{tr}A$ for any $2 \times 2$ matrix $A$.
We define the universal dressed K-operators for Q-operators by
\begin{align}
{\mathcal K}^{(a)}(x)&={\mathcal L}^{(a)}(x^{-1}) (\Kf^{(a)}(x)\otimes 1) 
\overline{\mathcal L}^{(a)}(x), 
\qquad x \in \mathbb{C}^{\times},
 \quad a=1,2.
 \label{dressed-UK-Q}
\end{align}
One can prove  
that \eqref{dressed-UK-Q} satisfy 
the following universal  dressed reflection equations 
for Q-operators 
 (see Appendix C).
\begin{multline}
\Lf^{(a)}_{12} \left(x^{-1} y \right){\mathcal K}^{(a)}_{13}(x) 
\overline{\Lf}^{(a)}_{12} \left( xy \right) 
{\mathcal K}_{23}(y) 
={\mathcal K}_{23}(y)
 \Lf^{(a)}_{12} \left(x^{-1}y^{-1} \right) {\mathcal K}^{(a)}_{13}(x) 
 \overline{\Lf}^{(a)}_{12} \left(x y^{-1} \right),
 \\
 x,y \in \mathbb{C}^{\times},
 \quad a=1,2.
\label{RE-dressL-Qti}
\end{multline}
We define the 
universal Q-operators \cite{BT18} by
\begin{align}
{\mathcal Q}^{(a)} (x) & = 
(\mathrm{tr}_{W_{a}} \otimes 1)
\left(
\check{\overline{\mathbf K}}^{(a)}(x^{-1}) 
{\mathcal K}^{(a)}(x) 
\right) , 
 \qquad x \in \mathbb{C}^{\times},
\quad  a =1,2,
\label{Q-opti}
\end{align}
where $\g^{(a)}=q^{\frac{(s_{0}-s_{1})\hs_{a}}{s}}$ and 
$W_{a}$ are Fock spaces generated by $\mathrm{Osc}_{a}$. 
Note that the second Q-operator follows from the first one 
\begin{align}
{\mathcal Q}^{(2)} (x)=\zeta \circ \sigma ({\mathcal Q}^{(1)} (x)).
\label{Q1toQ2}
\end{align}
One can show this by using the relations \eqref{UL2}, \eqref{ULb1} and \eqref{Kop3}. 
We find that the universal T-and Q-operators satisfy the following universal TQ-relations 
(see Appendix B for derivation).
\begin{multline}
(q^{2}-q^{4}x^{2s})
{\mathcal Q}^{(a)}(q^{\frac{1}{s}}x)  \mathcal{T}(x)  = 
\omega_{1}^{(a)}(x) \overline{\omega}_{1}^{(a)}(x) {\mathcal Q}^{(a)}(q^{-\frac{1}{s}}x)q^{h_{2-a}}
\\
 +
\omega_{2}^{(a)}(x) \overline{\omega}_{2}^{(a)}(x){\mathcal Q}^{(a)}(q^{\frac{3}{s}}x)q^{-h_{2-a}} ,  
\qquad 
a=1,2,
\label{TQ-uni}
\end{multline}
where 
$\omega^{(a)}_{1}(x), \omega^{(a)}_{2}(x),  \overline{\omega}^{(a)}_{1}(x), \overline{\omega}^{(a)}_{2}(x)$ are defined by 
\begin{align}
\begin{split}
\omega^{(1)}_{1}(x)
&=(\epsilon_{+}x^{s_{0}}+\epsilon_{-}x^{-s_{1}}) ,
\qquad 
\omega^{(1)}_{2}(x)=
 (1-x^{-2s})(\epsilon_{+}x^{-s_{0}}+\epsilon_{-}q^{2}x^{s_{1}})  q^{-2},
\\
\overline{\omega}^{(1)}_{1}(x)
&=(1-x^{2s}q^{4})(\overline{\epsilon}_{+}x^{-s_{0}}+\overline{\epsilon}_{-}x^{s_{1}})  q^{-1},
\quad
\overline{\omega}^{(1)}_{2}(x)=
 -x^{2s}(\overline{\epsilon}_{+}x^{s_{0}}+\overline{\epsilon}_{-}q^{-2}x^{-s_{1}}) q^{7},
  \end{split}
\label{omega}
\end{align} 
 and 
\begin{align}
 \omega^{(2)}_{j}(x)=\zeta(\omega^{(1)}_{j}(x)),   
 \qquad 
 \overline{\omega}^{(2)}_{j}(x)=\zeta(\overline{\omega}^{(1)}_{j}(x)), 
 \qquad 
 j=1,2. 
\end{align}
The universal T-and Q-operators commute with any Cartan elements of $U_{q}(\widehat{sl}_{2})$. 
One can show this by the relation $(1 \otimes q^{\xi h_{i}}){\mathcal R}=(q^{-\xi h_{i}} \otimes 1)
{\mathcal R}(q^{\xi h_{i}} \otimes q^{\xi h_{i}})$ 
from \eqref{uniCar}, the fact that Cartan elements commute with the K-operators, and cyclicity of the trace. 
Thus one can remove the factors $q^{h_{2-a}}$ and  $q^{-h_{2-a}}$ in \eqref{TQ-uni} 
by setting ${\mathcal Q}^{(a)}(x)= x^{\frac{s h_{2-a}}{2}}{\mathcal Q}^{(a)\prime}(x)$. 
The universal T-operator \eqref{T-op-fin} and Q-operators \eqref{Q-opti} belong to $U_{q}(\widehat{sl}_{2})$. 
Thus the universal TQ-relations \eqref{TQ-uni} are equations in $U_{q}(\widehat{sl}_{2})$. 
Evaluating these for various representations of  $U_{q}(\widehat{sl}_{2})$, 
we obtain a wide class of  T-and Q-operators. 
For example, the T-and Q-operators acting on 
$({\mathbb C}^{2})^{\otimes L}$ are given by (see eqs. (G.22) and (G.24) in \cite{BT18})
\footnote{Note that the following relations follow from \eqref{R-def}: 
$(1 \otimes \Delta^{\otimes (L-1)}) \Rc =\Rc_{0L}\cdots \Rc_{02} \Rc_{01}$, 
$(1 \otimes \Delta^{\otimes (L-1)}) \overline{\Rc} =\overline{\Rc}_{01}\overline{\Rc}_{02} \cdots  \overline{\Rc}_{0L}$. 
Here we label the space over which the trace is to be taken (auxiliary space) as $0$. 
One can also consider higher spin representations in the quantum space similarly. 
In particular for finite dimensional representations, the universal L-operators reduce to finite size L-operators  
since the generators $E$ and $F$ become nilpotent.}
\begin{multline}
{\mathbf T}(x) =\Psi(x,\{\xi_{i}\}) \left( \pi_{\xi_{1}} \otimes \dots \otimes \pi_{\xi_{L}} \right )
 \Delta^{\otimes (L-1)}  {\mathcal T}(x) ,
\\
=
(\mathrm{tr} \otimes 1^{\otimes L})
\Bigl(
\overline{K}_{0}
\left(x^{-1}\right)  
R_{0L}\left(x^{-1}\xi_{L}^{-1}\right)
\cdots 
R_{01}\left(x^{-1}\xi_{1}^{-1}\right)
\\
\times 
K_{0}(x) 
\overline{R}_{01}\left(x \xi_{1}^{-1}\right)
\cdots 
\overline{R}_{0L}\left(x \xi_{L}^{-1} \right)
\Bigr), 
\label{T-op-latti}
\end{multline}
\begin{multline}
{\mathbf Q}^{(a)}(x) = \Psi^{(1)}(x,\{\xi_{i}\})
\left( \pi_{\xi_{1}} \otimes \dots \otimes \pi_{\xi_{L}} \right )
 \Delta^{\otimes (L-1)} {\mathcal{Q}}^{(a)}(x) ,
\\
=
(\mathrm{tr}_{W_{a}} \otimes 1^{\otimes L})
\Bigl(
\check{\overline{\mathbf K}}^{(a)}_{0} 
\left(x^{-1} \right)
{\mathbf L}^{(a)}_{0L}\left(x^{-1} \xi_{L}^{-1} \right)
\cdots 
{\mathbf L}^{(a)}_{01}\left(x^{-1} \xi_{1}^{-1} \right)
\\
\times 
{\mathbf K}^{(a)}_{0} (x)
\overline{\mathbf L}^{(a)}_{01}\left(x \xi_{1}^{-1} \right)
\cdots 
\overline{\mathbf L}^{(a)}_{0L}\left(x \xi_{L}^{-1} \right)
\Bigr), 
\quad a=1,2, 
 \label{Q-op'-fun}
\end{multline}
where $\xi_{1},\dots ,\xi_{L} \in {\mathbb C}^{\times}$ are inhomogeneities on the spectral parameter 
in the quantum space.  
The overall factors
\footnote{$\Psi^{(1)}(x,\{\xi_{i}\})$ and $\Psi(x,\{\xi_{i}\})$ in \cite{BT18} correspond to 
$\Psi^{(1)}(x,\{\xi_{i}\})^{-1}$ and $q^{L}\Psi(x,\{\xi_{i}\})^{-1}$, respectively.}
 are given by 
\begin{align}
\Psi(x,\{\xi_{i}\})&=q^{L}\prod_{k=1}^{L} 
\phi(x^{-1} \xi^{-1}_{k}) \phi(x^{-1} \xi_{k}), 
\\
\Psi^{(1)}(x,\{\xi_{i}\})&=\prod_{k=1}^{L} 
\phi^{(1)}(x^{-1} \xi^{-1}_{k}) \phi^{(1)}(x^{-1} \xi_{k}), 
\qquad a=1,2.
\end{align}
We remark that  \eqref{Q-op'-fun} give Q-operators for the open XXZ-spin chain. 
Moreover, our Q-operators
\eqref{Q-op'-fun} 
reduce to Q-operators for the open XXX-spin chain similar to 
the ones in \cite{FS15} in the rational limit $q \to 1$. 
Applying  $\left( \pi_{\xi_{1}} \otimes \dots \otimes \pi_{\xi_{L}} \right )
 \Delta^{\otimes (L-1)}$ to \eqref{TQ-uni}, we obtain the Baxter TQ-relations for the open XXZ-spin chain. 
\begin{multline}
(q^{2}-q^{4}x^{2s})
{\mathbf Q}^{(a)}(q^{\frac{1}{s}}x)  \mathbf{T}(x)  = 
\omega^{(a)}_{1}(x)\overline{\omega}^{(a)}_{1}(x) \chi_{1}(x)  {\mathbf Q}^{(a)}(q^{-\frac{1}{s}}x) \eta^{3-2a}
\\
 +
\omega^{(a)}_{2}(x) \overline{\omega}^{(a)}_{2}(x) \chi_{2}(x) {\mathbf Q}^{(a)}(q^{\frac{3}{s}}x) \eta^{-3+2a},
\qquad a=1,2,
\label{TQ-XXZ}
\end{multline}
where 
$\omega^{(a)}_{1}(x), \omega^{(a)}_{2}(x),  \overline{\omega}^{(a)}_{1}(x)$ and $\overline{\omega}^{(a)}_{2}(x)$ are defined by \eqref{omega}, 
and $\chi_{1}(x)$ and $\chi_{2}(x)$ are calculated as
\begin{align}
\chi_{1}(x) &= \frac{\Psi^{(1)}(q^{\frac{1}{s}}x,\{\xi_{i}\})\Psi(x,\{\xi_{i}\})}{\Psi^{(1)}(q^{-\frac{1}{s}}x,\{\xi_{i}\})}
\nonumber 
\\
&=q^{L}\prod_{k=1}^{L}\frac{\phi^{(1)}(q^{-\frac{1}{s}}x^{-1} \xi^{-1}_{k}) \phi^{(1)}(q^{-\frac{1}{s}}x^{-1} \xi_{k})
\phi(x^{-1} \xi^{-1}_{k}) \phi(x^{-1} \xi_{k})}
{\phi^{(1)}(q^{\frac{1}{s}}x^{-1} \xi^{-1}_{k}) \phi^{(1)}(q^{\frac{1}{s}}x^{-1} \xi_{k})}
\nonumber 
\\
&=q^{L}\prod_{k=1}^{L} \left(1-q^{-2}(x  \xi_{k})^{-s}\right) \left(1-q^{-2}(x  \xi_{k}^{-1})^{-s}\right)  ,
\label{coef1}
\\
\chi_{2}(x) &= \frac{\Psi^{(1)}(q^{\frac{1}{s}}x,\{\xi_{i}\})\Psi(x,\{\xi_{i}\})}{\Psi^{(1)}(q^{\frac{3}{s}}x,\{\xi_{i}\})}
\nonumber 
\\
&=q^{L}\prod_{k=1}^{L}\frac{\phi^{(1)}(q^{-\frac{1}{s}}x^{-1} \xi^{-1}_{k}) \phi^{(1)}(q^{-\frac{1}{s}}x^{-1} \xi_{k})
\phi(x^{-1} \xi^{-1}_{k}) \phi(x^{-1} \xi_{k})}
{\phi^{(1)}(q^{-\frac{3}{s}}x^{-1} \xi^{-1}_{k}) \phi^{(1)}(q^{-\frac{3}{s}}x^{-1} \xi_{k})}
\nonumber 
\\
&=q^{L}\prod_{k=1}^{L} \left(1-(x  \xi_{k})^{-s}\right) \left(1-(x  \xi_{k}^{-1})^{-s}\right)  .
\label{coef2}
\end{align}
The factors $\eta$ and $\eta^{-1}$ are diagonal matrices.
\begin{align}
\eta =\begin{pmatrix}
q  & 0  \\
0  & q^{-1}
 \end{pmatrix}^{\otimes L}
,
\qquad 
\eta^{-1} =\begin{pmatrix}
q^{-1}  & 0  \\
0  & q
 \end{pmatrix}^{\otimes L}
 .
\end{align} 
Up to convention, \eqref{TQ-XXZ} for $a=1$ coincides
\footnote{Set $s_{0}=s_{1}=1$, $s=2$. In this case, $R(x)=\overline{R}(x^{-1})$ and $\Lf^{(1)}(x)=\overline{\Lf}^{(1)}(x^{-1})$ 
hold. Then we make identification: 
$\mathbf{T}(x)=q^{2L+1}\epsilon_{+} \overline{\epsilon}_{-}x^{2}{\mathcal T}^{V}(x^{-1})$, 
${\mathbf Q}^{(a)}(x)=(\overline{\epsilon}_{+}q^{-1}x^{-2}/\overline{\epsilon}_{-}) {\mathcal T}^{W}(q^{\frac{1}{2}}x^{-1})
=(\overline{\epsilon}_{+}q^{-1}x^{-2}/\overline{\epsilon}_{-}) \begin{pmatrix}q^{-1}x^{2} & 0 \\ 0 & 1 \end{pmatrix}^{\otimes L}{\mathcal Q}(q^{\frac{1}{2}}x^{-1})$, $\xi_{j}=t^{-1}_{j}$, 
$-\epsilon_{-}/\epsilon_{+}=\xi$,
$-\overline{\epsilon}_{+}/\overline{\epsilon}_{-}=\tilde{\xi}$, 
$L=N$,
$K(x)=- \epsilon_{+} x K^{V}(x^{-1})$,
$\overline{K}(x)=-q\overline{\epsilon}_{-}x^{-1}\widetilde{K}^{V}(x)$, 
$R(x)=q(R(x) \text{ in eq. (2.15) in \cite{VW20}})$, 
$\Lf^{(1)}(x)=L(q^{\frac{1}{2}}x,1)$, 
$\Kf^{(1)}(x)=K^{W}(q^{\frac{1}{2}}x^{-1},1)$, 
$\Kfcb^{(1)}(x)=(\overline{\epsilon}_{+}q^{-1}x^{2}/\overline{\epsilon}_{-})\widetilde{K}^{V}(x,1)$, 
$\es_{1}=-q^{-\frac{1}{2}}\lambda^{-1} a$, 
$\fs_{1}=-q^{\frac{3}{2}}\lambda^{-1} a^{\dagger}$, 
$\hs_{1}=-2D$, 
where the quantities in the left hand sides 
are in the notation of the present paper, and those 
 in the right hand sides are mainly expressed in the notation of \cite{VW20}. 
One has to replace $q$ with $q^{-1}$ and $x$ with $x^{-1}$, and 
reverse the ordering of the lattice sites to make comparison. 
}
 with the TQ-relation in \cite{VW20}.

 Commutativity of the universal T-and Q-operators can be shown 
 based on a refinement of the Sklyanin's method \cite{Skly} 
 (with the help of \eqref{LLcb1=c}-\eqref{LbLc1s=c}, \eqref{tt}, 
\eqref{refeqlim2ndch-pr} and \eqref{RE-dressL-Qti})
 as explained in Appendix G in \cite{BT18}: 
\begin{align}
\mathcal{Q}^{(a)}(x)\mathcal{T}(y)=
\mathcal{T}(y)\mathcal{Q}^{(a)}(x), 
\quad 
a=1,2, 
\ 
x,y \in {\mathbb C}. 
 \label{commutativity}
\end{align}
Evaluating \eqref{commutativity} for $\left( \pi_{\xi_{1}} \otimes \dots \otimes \pi_{\xi_{L}} \right )
 \Delta^{\otimes (L-1)}$, we also obtain
\begin{align} 
{\mathbf Q}^{(a)}(x){\mathbf T}(y)=
{\mathbf T}(y){\mathbf Q}^{(a)}(x) , 
\quad 
a=1,2, 
\ 
x,y \in {\mathbb C}. 
 \label{commutativityti}
\end{align}
\section{Concluding remarks}
In this paper, we gave universal  Baxter TQ-relations for diagonal open  boundary 
integrable systems associated with $U_{q}(\widehat{sl_{2}})$. 
This supplements and expands the discussions in \cite{BT18}. 
By fixing the representation on 
the quantum space, we 
recovered the Baxter TQ-relation for the open XXZ spin chain \cite{VW20}. 

One of the unsolved problems related to this paper is generalization to quantum integrable systems with 
non-diagonal open boundaries. The key objects for construction of the Baxter Q-operators 
for open boundary integrable systems are K-operators. 
The K-operators for Q-operators can be obtained as asymptotic limits (or contraction) of generic 
K-operators which are expressed in terms of generators of symmetry algebras. 
In the case of the Yangian $Y(sl_2)$,
generic K-operators for general non-diagonal boundaries were constructed in \cite{FGK19}, 
and in the case of  $U_{q}(\widehat{sl_{2}})$, 
generic K-operators for general triangular boundaries were constructed in  \cite{T19}. 

The generalization to the higher rank case is also not fully understood yet. 
In \cite{Ts18}, diagonal K-operators for $U_{q}(\widehat{gl_{n}})$ were expressed 
in terms of  Cartan elements of  a quotient of $U_{q}(gl_{n})$. 
 Non diagonal K-matrices for the symmetric tensor representations of $U_{q}(\widehat{sl_{n}})$ 
were constructed in  \cite{KOY18} (see also \cite{ML19} for some aspect on $n=2$ case). 
By taking limits of these, one will be able to obtain a subset of the K-operators for Q-operators 
for the higher rank case. 

Another way to construct Baxter TQ-relations for open boundaries would be to use 
a generating function of the T-operators for the anti-symmetric representations. 
In the case of the periodic boundary condition, it is 
 a column-ordered determinant over a function of the monodromy matrix for a transfer 
matrix \cite{Ta04}. For models with  open boundaries, 
one may have to use a dressed K-matrix 
(in our case, the universal dressed K-operator \eqref{dressed-UKti}) 
instead of the usual monodromy matrix. 

It is known that T-operators can be expressed as concise Wronskian-like  determinants (Casoratian)
\footnote{In this context, the QQ-relations are fundamental objects. 
In order to derive the QQ-relations, one would have to consider analogues of \eqref{GKLbKG} and \eqref{GKbLbKbG} among two 
K-operators in \eqref{Kop1}, \eqref{Kop2} and \eqref{Kop3}, 
in addition to analogues of  \eqref{GL1LGuni} and \eqref{GLb1LGuni} among two universal L-operators in \eqref{UL1}-\eqref{ULb1}
 (cf. (4.11) in \cite{KT14}).}
 in terms of Q-operators 
(in addition to the references for Baxter Q-operators referred in Introduction, see also \cite{KLWZ96,T09,T11,KLV15,ESV20} and references therein). 
In contrast with integrable systems with periodic boundary, not much is known about this for integrable systems with open boundaries (cf. \cite{Ne19}). 

\section*{Acknowledgments} 
The author would like to thank Pascal Baseilhac for collaboration in the previous paper \cite{BT18}, 
and Michio Jimbo for 
answering a question on  unitarity relations of R-matrices. 
He also thanks the anonymous referee for useful comments. 
The work is supported by Grant No. 0657-2020-0015 of the Ministry of Science and Higher Education of Russia. 
A part of this work was previously announced at the conference 
`Mini-workshop on Symmetry and Interactions', 
Shing-Tung Yau Center of Southeast University (Nanjing, China, 23 November 2019).
\section*{Appendix A: The universal R-matrix}
\label{ApA}
\addcontentsline{toc}{section}{Appendix A}
\def\theequation{A\arabic{equation}}
\setcounter{equation}{0}
In this section, we briefly review the product expression of the universal R-matrix given by 
Khoroshkin and Tolstoy in \cite{TK92,KT94}. 
Their universal R-matrix was already reviewed by several authors 
(see for example, \cite{ZG93,KST94,BGKNR10,BGKNR12,MT15,BT18,PSZ16}). 
Here we basically follow these in the convention in Appendix A in \cite{BT18}.

Let $\{\alpha +k\delta \}_{k=0}^{\infty} \cup
 \{k\delta \}_{k=1}^{\infty}\cup
  \{\delta- \alpha +k\delta \}_{k=0}^{\infty} $ be a positive root system of $\widehat{sl_2}$ 
in the notation of \cite{TK92}. We fix the root ordering as 
 $\alpha +(k-1)\delta \prec \alpha +k\delta \prec 
 l\delta \prec (l+1)\delta \prec 
 \delta- \alpha +m\delta \prec \delta- \alpha +(m-1)\delta$ for any $k,l,m \in {\mathbb Z}_{\ge 1}$. 
 In this case, 
the universal R-matrix has the following explicit expression:
\begin{align}
{\mathcal R}=\overline{\mathcal R}^{+} \, 
\overline{\mathcal R}^{0} \,
 \overline{\mathcal R}^{-}q^{\frac{1}{2} h_{1}\otimes h_{1}},  
  \label{UR-prod}
\end{align}
where each element is defined by 
\begin{align}
\begin{split}
\overline{\mathcal R}^{+}&=\overrightarrow{\prod_{k=0}^{\infty}} \exp_{q^{-2}} 
\left( \lambda e_{\alpha + k \delta } \otimes f_{\alpha + k \delta } \right) ,
\\[6pt]
\overline{\mathcal R}^{0}&= \exp  
\left( \lambda   \sum_{k=1}^{\infty} \frac{k}{[2k]_{q}} e_{ k \delta } \otimes f_{ k \delta } \right) ,
\\[6pt]
\overline{\mathcal R}^{-}&=\overleftarrow{\prod_{k=0}^{\infty}} \exp_{q^{-2}} 
\left( \lambda e_{\delta-\alpha + k \delta } \otimes f_{\delta-\alpha + k \delta } \right) .
\end{split}
\end{align}
Let 
$e_{\alpha } =e_{1}$, $e_{\delta- \alpha } =e_{0}$, 
$f_{\alpha } =f_{1}$, $f_{\delta- \alpha } =f_{0}$. Then the other root vectors are defined by 
the following recursion relations:
\begin{align}
\begin{split}
e_{\alpha +k \delta } &= [2]_{q}^{-1}[e_{\alpha +(k-1) \delta}, e_{\delta}^{\prime}], 
\\[6pt]
e_{k \delta }^{\prime } &= [e_{\alpha +(k-1) \delta}, e_{\delta -\alpha }]_{q^{-2}}, 
\\[6pt]
e_{\delta- \alpha +k \delta } &= [2]_{q}^{-1}[ e_{\delta}^{\prime}, e_{\delta- \alpha +(k-1) \delta}],
\\[6pt]
f_{\alpha +k \delta } &= [2]_{q}^{-1}[f_{\delta}^{\prime}, f_{\alpha +(k-1) \delta}], 
\\[6pt]
f_{k \delta }^{\prime } &= [f_{\delta -\alpha }, f_{\alpha +(k-1) \delta}]_{q^{2}}, 
\\[6pt]
f_{\delta- \alpha +k \delta } &= [2]_{q}^{-1}[ f_{\delta- \alpha +(k-1) \delta}, f_{\delta}^{\prime}],  
\qquad k \in {\mathbb Z}_{\ge 1} ,
\end{split}
\label{root1}
\end{align}
and the following generating functions:
\begin{align}
\begin{split}
\lambda \sum_{k=1}^{\infty} e_{k \delta} z^{-k}&= 
\log \left(  1+ \lambda \sum_{k=1}^{\infty} e_{k \delta}^{\prime } z^{-k} \right),
\\[6pt]
-\lambda \sum_{k=1}^{\infty} f_{k \delta} z^{-k}&= 
\log \left(  1- \lambda \sum_{k=1}^{\infty} f_{k \delta}^{\prime } z^{-k} \right), 
\qquad z \in {\mathbb C}.
\end{split}
\label{root2}
\end{align}
In general, root vectors contain many (q-deformed) commutators. 
However, simplification occurs under the 
evaluation map. 
For $ k \in {\mathbb Z}_{\ge 0}$, we have 
\begin{align}
\mathsf{ev}_{x}(e_{\alpha +k \delta }) &=(-1)^{k}x^{ks+s_{1}}q^{-kH}E,
 \label{afev1}
\\[6pt]
\mathsf{ev}_{x}(e_{\delta- \alpha +k \delta }) &=(-1)^{k}x^{ks+s_{0}}Fq^{-kH}, 
 \label{afev2}
\\[6pt]
\mathsf{ev}_{x}(f_{\alpha +k \delta }) &=(-1)^{k}x^{-ks-s_{1}}Fq^{kH}, 
 \label{afev3}
\\
\mathsf{ev}_{x}(f_{\delta- \alpha +k \delta }) &=(-1)^{k}x^{-ks-s_{0}}q^{kH}E ,
 \label{afev4}
\end{align}
and for $k \in {\mathbb Z}_{\ge 1}$, 
\begin{align}
\mathsf{ev}_{x}(e_{k \delta }^{\prime}) 
%
&=(-1)^{k-1}x^{ks}q^{-(k-1)H-k}\left(\lambda [k]_{q}C-\frac{[k-1]_{q}q^{H}+[k+1]_{q}q^{-H}}{\lambda} \right), 
 \label{afev5}
\\
\mathsf{ev}_{x}(e_{k \delta }) &=
\frac{(-1)^{k-1}q^{-k}x^{ks}}{(q-q^{-1})k}\left(C_{k}-(q^{k}+q^{-k})q^{-kH}\right), 
 \label{afev6}
\\
\mathsf{ev}_{x}(f_{k \delta }^{\prime}) 
%
&
=(-1)^{k-1}x^{-ks}q^{(k-1)H+k}
\left(-\lambda [k]_{q}C+\frac{[k+1]_{q}q^{H}+[k-1]_{q}q^{-H}}{\lambda} \right),
 \label{afev7}
\\
\mathsf{ev}_{x}(f_{k \delta }) &=-
\frac{(-1)^{k-1}q^{k}x^{-ks}}{(q-q^{-1})k}\left(C_{k}-(q^{k}+q^{-k})q^{kH}\right),
 \label{afev8}
\end{align}
where the central elements $C_{k}$ are defined by 
\begin{align}
\sum_{k=1}^{\infty}\frac{(-1)^{k-1}C_{k}}{k}z^{-k}= \log(1+\lambda^{2} C z^{-1}+z^{-2}), 
\qquad z \in {\mathbb C}.
\label{highercas}
\end{align}
%
In addition to the root vectors defined above, we will also use another set of root vectors. 
Let 
$\overline{e}_{\alpha } =e_{\alpha}$, $\overline{e}_{\delta- \alpha } =e_{\delta- \alpha}$, 
$\overline{f}_{\alpha } =f_{\alpha}$, $\overline{f}_{\delta- \alpha } =f_{\delta- \alpha}$. Then the other root vectors are defined by 
the following recursion relations:
\begin{align}
\begin{split}
\overline{e}_{\alpha +k \delta } &= [2]_{q}^{-1}[\overline{e}_{\alpha +(k-1) \delta}, \overline{e}_{\delta}^{\prime}], 
\\[6pt]
\overline{e}_{k \delta }^{\prime } &= [e_{\alpha}, \overline{e}_{\delta -\alpha +(k-1) \delta }]_{q^{2}}, 
\\[6pt]
\overline{e}_{\delta- \alpha +k \delta } &= [2]_{q}^{-1}[ \overline{e}_{\delta}^{\prime}, \overline{e}_{\delta- \alpha +(k-1) \delta}],
\\[6pt]
\overline{f}_{\alpha +k \delta } &= [2]_{q}^{-1}[\overline{f}_{\delta}^{\prime}, \overline{f}_{\alpha +(k-1) \delta}], 
\\[6pt]
\overline{f}_{k \delta }^{\prime } &= [\overline{f}_{\delta -\alpha +(k-1) \delta}, f_{\alpha }]_{q^{-2}}, 
\\[6pt]
\overline{f}_{\delta- \alpha +k \delta } &= [2]_{q}^{-1}[\overline{f}_{\delta- \alpha +(k-1) \delta}, \overline{f}_{\delta}^{\prime}],  
\qquad k \in {\mathbb Z}_{\ge 1} ,
\end{split}
\label{root3}
\end{align}
and  the following generating functions: 
\begin{align}
\begin{split}
-\lambda \sum_{k=1}^{\infty} q^{-2k}\overline{e}_{k \delta} z^{-k}&= 
\log \left(  1- \lambda \sum_{k=1}^{\infty} q^{-2k}\overline{e}_{k \delta}^{\prime } z^{-k} \right),
\\[6pt]
\lambda \sum_{k=1}^{\infty} q^{2k}\overline{f}_{k \delta} z^{-k}&= 
\log \left(  1+ \lambda \sum_{k=1}^{\infty} q^{2k}\overline{f}_{k \delta}^{\prime } z^{-k} \right), 
\qquad z \in {\mathbb C}.
\end{split}
\label{root4}
\end{align}
One can prove the following relations by induction.
\begin{align}
\begin{split}
& \sigma(e_{\alpha +k\delta})=q^{-2k}\overline{e}_{\delta-\alpha +k\delta}, 
\quad 
\sigma(e_{\delta-\alpha +k\delta})=q^{-2k}\overline{e}_{\alpha +k\delta}, 
\\
& \sigma(f_{\alpha +k\delta})=q^{2k}\overline{f}_{\delta-\alpha +k\delta}, 
\quad 
\sigma(f_{\delta-\alpha +k\delta})=q^{2k}\overline{f}_{\alpha +k\delta}
\quad \text{for} \quad k \in {\mathbb Z}_{\ge 0} ,
\\
& \sigma(e^{\prime}_{k\delta})=-q^{-2k}\overline{e}^{\prime}_{k\delta}, 
\quad
\sigma(f^{\prime}_{k\delta})=-q^{2k}\overline{f}^{\prime}_{k\delta} ,
\\
& \sigma(e_{k\delta})=-q^{-2k}\overline{e}_{k\delta}, 
\quad
\sigma(f_{k\delta})=-q^{2k}\overline{f}_{k\delta}
\quad \text{for} \quad k \in {\mathbb Z}_{\ge 1}.
\end{split}
 \label{sigroots}
\end{align}
\section*{Appendix B: Derivation of the universal Baxter TQ-relations}
\label{ApB}
\addcontentsline{toc}{section}{Appendix B}
\def\theequation{B\arabic{equation}}
\setcounter{equation}{0}
In this section, we derive  the universal Baxter TQ-relations \eqref{TQ-uni} under the assumption
\footnote{This assumption was already verified for concrete models in \cite{FS15,VW20}, but remains 
an open problem on the level of the universal Baxter Q-operators.}
 that 
 convergence and cyclicity of the traces in the T-and Q-operators hold. 
 Baxter TQ-relations with fixed quantum spaces were already derived for the open XXX-spin chains \cite{FS15} and 
 for the open XXZ-spin chains \cite{VW20}. Here we consider the problem on the level of 
 the universal T-and Q-operators. 

We introduce two kinds of elements $\G, \Gb \in \mathrm{Osc}_{1}\otimes \mathrm{End}(\mathbb{C}^{2})$ and their inverse:
\begin{align}
\G&=q^{-\frac{\hs_{1}}{2}} \otimes E_{11}+q^{\frac{\hs_{1}}{2}} \otimes E_{22}- 
\lambda q^{-\frac{\hs_{1}}{2}-\frac{s_{0}}{s}} \fs_{1} \otimes E_{12}  , 
\label{G}
\\
\G^{-1}&=q^{\frac{\hs_{1}}{2}} \otimes E_{11}+q^{-\frac{\hs_{1}}{2}} \otimes E_{22}+ 
\lambda q^{-\frac{\hs_{1}}{2}-\frac{s_{0}}{s}-1} \fs_{1} \otimes E_{12}  ,
\label{Gin}
\\
\Gb&=q^{-\frac{\hs_{1}}{2}} \otimes E_{11}+q^{\frac{\hs_{1}}{2}} \otimes E_{22}- 
\lambda q^{-\frac{\hs_{1}}{2}-\frac{s_{1}}{s}} \fs_{1} \otimes E_{12}  , 
\label{Gb}
\\
\Gb^{-1}&=q^{\frac{\hs_{1}}{2}} \otimes E_{11}+q^{-\frac{\hs_{1}}{2}} \otimes E_{22}+ 
\lambda q^{-\frac{\hs_{1}}{2}-\frac{s_{1}}{s}-1} \fs_{1} \otimes E_{12}  .
\label{Gin}
\end{align}
One can check the following relations by direct calculations.
\begin{align}
& \G^{-1} (\rho^{(1)}_{xq^{-\frac{1}{s}}} \otimes \pi_{1})\Delta(q^{\xi h_{0}}) \G=
 q^{- \xi (\hs_{1}+1)} \otimes E_{11}+q^{-\xi (\hs_{1}-1)} \otimes E_{22}
  \nonumber 
\\
& \quad =
(q^{\frac{s_{0}-s_{1}}{2s} \hs_{1}}\rho^{(1)}_{xq^{\frac{1}{s}}}(q^{\xi (h_{0}-1)}) q^{-\frac{s_{0}-s_{1}}{2s} \hs_{1}} )\otimes E_{11}
+(q^{-\frac{s_{0}-s_{1}}{2s} \hs_{1}}\rho^{(1)}_{xq^{-\frac{3}{s}}}(q^{\xi (h_{0}+1)}) q^{\frac{s_{0}-s_{1}}{2s} \hs_{1}} )\otimes E_{22}
   ,  \label{rh1pih0}  
\\
& \G^{-1} (\rho^{(1)}_{xq^{-\frac{1}{s}}} \otimes \pi_{1})\Delta(q^{\xi h_{1}}) \G=
 q^{\xi (\hs_{1}+1)} \otimes E_{11}+ q^{\xi (\hs_{1}-1)} \otimes E_{22}
  \nonumber 
\\
& \quad =
(q^{\frac{s_{0}-s_{1}}{2s} \hs_{1}}\rho^{(1)}_{xq^{\frac{1}{s}}}(q^{\xi (h_{1}+1)}) q^{-\frac{s_{0}-s_{1}}{2s} \hs_{1}} )\otimes E_{11}
+(q^{-\frac{s_{0}-s_{1}}{2s} \hs_{1}}\rho^{(1)}_{xq^{-\frac{3}{s}}}(q^{\xi (h_{1}-1)}) q^{\frac{s_{0}-s_{1}}{2s} \hs_{1}} )\otimes E_{22}
   ,  \label{rh1pih1}  
\\
& \G^{-1} (\rho^{(1)}_{xq^{-\frac{1}{s}}} \otimes \pi_{1})\Delta(e_{0}) \G=
  x^{s_{0}}q^{1-\frac{s_{0}}{s}}\fs_{1} \otimes E_{11}+x^{s_{0}}q^{-1-\frac{s_{0}}{s}}\fs_{1} \otimes E_{22}+
x^{s_{0}} \otimes E_{21}
\nonumber 
\\
& \quad =
(q^{\frac{s_{0}-s_{1}}{2s} \hs_{1}}\rho^{(1)}_{xq^{\frac{1}{s}}}(e_{0}) q^{-\frac{s_{0}-s_{1}}{2s} \hs_{1}} )\otimes E_{11}
+(q^{-\frac{s_{0}-s_{1}}{2s} \hs_{1}}\rho^{(1)}_{xq^{-\frac{3}{s}}}(e_{0}) q^{\frac{s_{0}-s_{1}}{2s} \hs_{1}} )\otimes E_{22}+
x^{s_{0}} \otimes E_{21}   ,  \label{rh1pie0}  
\\
& \G^{-1} (\rho^{(1)}_{xq^{-\frac{1}{s}}} \otimes \pi_{1})\Delta(e_{1}) \G=
  x^{s_{1}}q^{1-\frac{s_{1}}{s}}\es_{1} \otimes E_{11}+x^{s_{1}}q^{-1-\frac{s_{1}}{s}}\es_{1} \otimes E_{22}
\nonumber 
\\
& \quad =
(q^{\frac{s_{0}-s_{1}}{2s} \hs_{1}}\rho^{(1)}_{xq^{\frac{1}{s}}}(e_{1}) q^{-\frac{s_{0}-s_{1}}{2s} \hs_{1}} )\otimes E_{11}
+(q^{-\frac{s_{0}-s_{1}}{2s} \hs_{1}}\rho^{(1)}_{xq^{-\frac{3}{s}}}(e_{1}) q^{\frac{s_{0}-s_{1}}{2s} \hs_{1}} )\otimes E_{22}   
,  \label{rh1pie1}  
\\
& \Gb^{-1} (\rho^{(1)}_{xq^{\frac{1}{s}}} \otimes \pi_{1})\Delta^{\prime }(q^{\xi h_{0}}) \Gb=
  q^{- \xi (\hs_{1}+1)} \otimes E_{11}+q^{- \xi (\hs_{1}-1)} \otimes E_{22}
\nonumber 
\\
& \quad =
(q^{\frac{s_{1}-s_{0}}{2s} \hs_{1}}\rho^{(1)}_{xq^{-\frac{1}{s}}}(q^{\xi (h_{0}-1)}) q^{-\frac{s_{1}-s_{0}}{2s} \hs_{1}} )\otimes E_{11}
+(q^{-\frac{s_{1}-s_{0}}{2s} \hs_{1}}\rho^{(1)}_{xq^{\frac{3}{s}}}(q^{\xi (h_{0}+1)}) q^{\frac{s_{1}-s_{0}}{2s} \hs_{1}} )\otimes E_{22} ,  \label{rh1ph0} 
\\
& \Gb^{-1} (\rho^{(1)}_{xq^{\frac{1}{s}}} \otimes \pi_{1})\Delta^{\prime }(q^{\xi h_{1}}) \Gb=
 q^{\xi (\hs_{1}+1)} \otimes E_{11}+ q^{\xi (\hs_{1}-1)} \otimes E_{22}
\nonumber 
\\
& \quad =
(q^{\frac{s_{1}-s_{0}}{2s} \hs_{1}}\rho^{(1)}_{xq^{-\frac{1}{s}}}(q^{\xi (h_{1}+1)}) q^{-\frac{s_{1}-s_{0}}{2s} \hs_{1}} )\otimes E_{11}
+(q^{-\frac{s_{1}-s_{0}}{2s} \hs_{1}}\rho^{(1)}_{xq^{\frac{3}{s}}}(q^{\xi (h_{1}-1)}z) q^{\frac{s_{1}-s_{0}}{2s} \hs_{1}} )\otimes E_{22} ,  \label{rh1ph1} 
\\
& \Gb^{-1} (\rho^{(1)}_{xq^{\frac{1}{s}}} \otimes \pi_{1})\Delta^{\prime }(f_{0}) \Gb=
  x^{-s_{1}}q^{1-\frac{s_{0}}{s}}\es_{1} \otimes E_{11}+x^{-s_{0}}q^{-1-\frac{s_{0}}{s}}\es_{1} \otimes E_{22}
\nonumber 
\\
& \quad =
(q^{\frac{s_{1}-s_{0}}{2s} \hs_{1}}\rho^{(1)}_{xq^{-\frac{1}{s}}}(f_{0}) q^{-\frac{s_{1}-s_{0}}{2s} \hs_{1}} )\otimes E_{11}
+(q^{-\frac{s_{1}-s_{0}}{2s} \hs_{1}}\rho^{(1)}_{xq^{\frac{3}{s}}}(f_{0}) q^{\frac{s_{1}-s_{0}}{2s} \hs_{1}} )\otimes E_{22} ,  \label{rh1pf0} 
\\
& \Gb^{-1} (\rho^{(1)}_{xq^{\frac{1}{s}}} \otimes \pi_{1})\Delta^{\prime }(f_{1}) \Gb=
  x^{-s_{1}}q^{1-\frac{s_{1}}{s}}\fs_{1} \otimes E_{11}+x^{s_{1}}q^{-1-\frac{s_{1}}{s}}\fs_{1} \otimes E_{22}+
x^{-s_{1}} \otimes E_{21}
\nonumber 
\\
& \quad =
(q^{\frac{s_{1}-s_{0}}{2s} \hs_{1}}\rho^{(1)}_{xq^{-\frac{1}{s}}}(f_{1}) q^{-\frac{s_{1}-s_{0}}{2s} \hs_{1}} )\otimes E_{11}
+(q^{-\frac{s_{1}-s_{0}}{2s} \hs_{1}}\rho^{(1)}_{xq^{\frac{3}{s}}}(f_{1}) q^{\frac{s_{1}-s_{0}}{2s} \hs_{1}} )\otimes E_{22}+
x^{-s_{1}} \otimes E_{21} , \label{rh1pf1}  
\end{align}
where $x \in \mathbb{C}$.
Let us apply \eqref{rh1pih0}-\eqref{rh1pf1} to the second equation in \eqref{R-def}. 
Taking note on the fact that the co-multiplication of the universal R-matrix 
has the form
\footnote{See also similar discussions in section 4 in \cite{KT14}.} 
$(\Delta \otimes 1){\mathcal R}=
\tilde{\mathcal R}(\{ \Delta(e_{0}) \otimes 1, \Delta(e_{1}) \otimes 1 ,
 1\otimes 1 \otimes f_{0}, 1\otimes 1 \otimes f_{1} \})
q^{ \frac{ \Delta( h_{1} )\otimes h_{1} }{2}}$, 
and the relations $E_{ij}E_{kl}=\delta_{jk}E_{il}$, we obtain
\begin{multline}
\G_{12}^{-1}\Lc^{(1)}_{13}(xq^{-\frac{1}{s}}) \Lc_{23}(x)\G_{12}
=
\\
= (q^{\frac{s_{0}-s_{1}}{2s}\hs_{1}} \otimes 1 \otimes 1) \Lc^{(1)}_{13}(xq^{\frac{1}{s}})
(q^{-\frac{s_{0}-s_{1}}{2s}\hs_{1}} \otimes E_{11} \otimes q^{\frac{1}{2}h_{1}})
\\
+
 (q^{-\frac{s_{0}-s_{1}}{2s}\hs_{1}} \otimes 1 \otimes 1) \Lc^{(1)}_{13}(xq^{-\frac{3}{s}})
(q^{\frac{s_{0}-s_{1}}{2s}\hs_{1}} \otimes E_{22} \otimes q^{-\frac{1}{2}h_{1}})
\\
+ \lambda x^{s_{0}} {\mathcal F}_{13}(x) (1 \otimes E_{21} \otimes q^{\frac{1}{2}h_{1}}),
\label{GL1LGuni}
\end{multline}
\begin{multline}
\Gb_{12}^{-1}\Lcb^{(1)}_{13}(xq^{\frac{1}{s}}) \Lcb_{23}(x)\Gb_{12}
=
\\
= (q^{\frac{s_{1}-s_{0}}{2s}\hs_{1}} \otimes 1 \otimes 1) \Lcb^{(1)}_{13}(xq^{-\frac{1}{s}})
(q^{-\frac{s_{1}-s_{0}}{2s}\hs_{1}} \otimes E_{11} \otimes q^{\frac{1}{2}h_{1}})
\\
+
 (q^{-\frac{s_{1}-s_{0}}{2s}\hs_{1}} \otimes 1 \otimes 1) \Lcb^{(1)}_{13}(xq^{\frac{3}{s}})
(q^{\frac{s_{1}-s_{0}}{2s}\hs_{1}} \otimes E_{22} \otimes q^{-\frac{1}{2}h_{1}})
\\
+
 x^{-s_{1}} \lambda \overline{\mathcal F}_{13}(x)
(1 \otimes E_{21} \otimes q^{\frac{1}{2}h_{1}}), 
\label{GLb1LGuni}
\end{multline}
where ${\mathcal F}_{13}(x)$ and $\overline{\mathcal F}_{13}(x)$ are 
elements in  $\mathrm{Osc}_{1}\otimes  \mathrm{End}({\mathbb C}^{2})  \otimes \mathcal{B}_{-}$ and
 $\mathrm{Osc}_{1}\otimes  \mathrm{End}({\mathbb C}^{2})  \otimes \mathcal{B}_{+}$, respectively.
\footnote{
Although explicit expressions of them are not necessary for the proof of the universal Baxter TQ-relation, 
one can calculate them  based on the explicit expression of the universal R-matrix. 
For example, we obtain
\begin{multline}
{\mathcal F}_{13}(x)=
  (q^{-\frac{s_{0}-s_{1}}{2s}\hs_{1}} \otimes 1 \otimes 1) \Lc^{(1)}_{13}(xq^{-\frac{3}{s}})
(q^{\frac{s_{0}-s_{1}}{2s}\hs_{1}} \otimes 1 \otimes 1)
\\
\times 
q^{-\frac{1}{2}\hs_{1}\otimes 1 \otimes h_{1}}
 \exp_{q^{-2}}^{-1}(\lambda x^{s_{0}}q^{-1-\frac{s_{0}}{s}}\fs_{1} \otimes 1 \otimes f_{0})
 \left(1 \otimes 1 \otimes 
 \sum_{k=0}^{\infty}(-q^{-1}x^{s})^{k}f_{\delta - \alpha +k\delta}
 \right)
 \\
 \times 
 \exp_{q^{-2}}(\lambda x^{s_{0}}q^{1-\frac{s_{0}}{s}}\fs_{1} \otimes 1 \otimes f_{0})
 q^{\frac{1}{2}\hs_{1}\otimes 1 \otimes h_{1}}.
 \end{multline}
%
Evaluating \eqref{GL1LGuni} and \eqref{GLb1LGuni} in the fundamental evaluation representation 
in the third component of the tensor product, we obtain
\begin{multline}
\G_{12}^{-1}\Lf^{(1)}_{13}(xq^{-\frac{1}{s}}) R_{23}(x)\G_{12}
=
\\
=(q-q^{-1}x^{s}) (q^{\frac{s_{0}-s_{1}}{2s}\hs_{1}} \otimes 1 \otimes 1) \Lf^{(1)}_{13}(xq^{\frac{1}{s}})
(q^{-\frac{s_{0}-s_{1}}{2s}\hs_{1}} \otimes E_{11} \otimes \pi(q^{\frac{H-1}{2}}))
\\
+
(1-x^{s}) (q^{-\frac{s_{0}-s_{1}}{2s}\hs_{1}} \otimes 1 \otimes 1) \Lf^{(1)}_{13}(xq^{-\frac{3}{s}})
(q^{\frac{s_{0}-s_{1}}{2s}\hs_{1}} \otimes E_{22} \otimes \pi(q^{-\frac{H-1}{2}}))
\\
+
\lambda x^{s_{0}} (q^{-\frac{s_{0}-s_{1}}{2s}\hs_{1}} \otimes 1 \otimes 1) \Lf^{(1)}_{13}(xq^{-\frac{3}{s}})
(q^{\frac{s_{0}-s_{1}}{2s}\hs_{1}-\hs_{1}} \otimes E_{21} \otimes E_{12}),
\label{GL1LG}
\end{multline}
\begin{multline}
\Gb_{12}^{-1}\Lfb^{(1)}_{13}(xq^{\frac{1}{s}}) \overline{R}_{23}(x)\Gb_{12}
=
\\
=(q-q^{-1}x^{-s}) (q^{\frac{s_{1}-s_{0}}{2s}\hs_{1}} \otimes 1 \otimes 1) \Lfb^{(1)}_{13}(xq^{-\frac{1}{s}})
(q^{-\frac{s_{1}-s_{0}}{2s}\hs_{1}} \otimes E_{11} \otimes \pi(q^{\frac{H-1}{2}}))
\\
+
(1-x^{-s}) (q^{-\frac{s_{1}-s_{0}}{2s}\hs_{1}} \otimes 1 \otimes 1) \Lfb^{(1)}_{13}(xq^{\frac{3}{s}})
(q^{\frac{s_{1}-s_{0}}{2s}\hs_{1}} \otimes E_{22} \otimes \pi(q^{-\frac{H-1}{2}}))
\\
+
\lambda x^{-s_{1}} (q^{-\frac{s_{1}-s_{0}}{2s}\hs_{1}} \otimes 1 \otimes 1) \Lfb^{(1)}_{13}(xq^{\frac{3}{s}})
(q^{\frac{s_{1}-s_{0}}{2s}\hs_{1}-\hs_{1}} \otimes E_{21} \otimes E_{12}),
\label{GLb1LbG}
\end{multline}
This type of relations \eqref{GL1LG} and \eqref{GLb1LbG} are known in \cite{BJMST06}.
}
%
One can also show the following relations by direct calculations.
\begin{multline}
\G_{12}^{-1}\Kf^{(1)}_{1}(xq^{\frac{1}{s}})\Lfb^{(1)}_{12}(x^2q^{\frac{1}{s}}) K_{2}(x)\Gb_{12}
=\omega^{(1)}_{1}(x) q^{(\frac{2s_{0}}{s}-\frac{1}{2})\hs_{1}} \Kf^{(1)}(xq^{-\frac{1}{s}})
 \otimes E_{11} 
 \\
 +
 \omega^{(1)}_{2}(x) q^{-(\frac{2s_{0}}{s}-\frac{1}{2})\hs_{1}} \Kf^{(1)}(xq^{\frac{3}{s}})
 \otimes E_{22} 
 +
 \omega^{(1)}_{21}(x) q^{-\frac{1}{2}\hs_{1}} \Kf^{(1)}(xq^{\frac{1}{s}})
 \es_{1}
 \otimes E_{21} ,
\label{GKLbKG}
\end{multline}
\begin{multline}
\G^{t_{1}t_{2}}_{12}\Kfcb^{(1)}_{1}(x^{-1}q^{-\frac{1}{s}})g_{2}\Lfcb^{(1)}_{12}(x^{-2}q^{-\frac{5}{s}})
g^{-1}_{2} \overline{K}^{t_{2}}_{2}(x^{-1})(\Gb^{-1}_{12})^{t_{1}t_{2}}
=
\\
=\overline{\omega}^{(1)}_{1}(x)
q^{-(\frac{2s_{0}}{s}-\frac{1}{2})\hs_{1}} \Kfcb^{(1)}(x^{-1}q^{\frac{1}{s}})
 \otimes E_{11} 
 %
 +\overline{\omega}^{(1)}_{2}(x)
  q^{(\frac{2s_{0}}{s}-\frac{1}{2})\hs_{1}} \Kfcb^{(1)}(x^{-1}q^{-\frac{3}{s}})
 \otimes E_{22} 
 \\
 +\overline{\omega}^{(1)}_{12}(x)
  q^{-\frac{3}{2}\hs_{1}} \Kfcb^{(1)}(x^{-1}q^{-\frac{1}{s}})
 \fs_{1}
 \otimes E_{12} ,
\label{GKbLbKbG}
\end{multline}
where the coefficients are defined by \eqref{omega} and 
\begin{align}
\begin{split}
 \omega^{(1)}_{21}(x)&=
 \lambda x^{-s}(\epsilon_{+}x^{s_{1}}+\epsilon_{-}x^{-s_{0}})  q^{-\frac{s_{0}}{s}}  ,
\\
\overline{\omega}^{(1)}_{12}(x)&=
 \lambda x^{2s_{1}}(\overline{\epsilon}_{+}qx^{s_{0}}+\overline{\epsilon}_{-}q^{-1}x^{-s_{1}}) 
  q^{\frac{s_{1}}{s}}  .
  \end{split}
\end{align} 
We remark that rational analogues of \eqref{GKLbKG} and \eqref{GKbLbKbG} were previously 
considered in \cite{FS15}. Moreover, the diagonal parts of \eqref{GKLbKG} and \eqref{GKbLbKbG}, 
which are essential in the proof of Baxter TQ-relations, appeared in \cite{VW20}. 
Define permutation operators by 
\begin{multline}
\mathfrak{p}_{12}(X\otimes Y \otimes Z)= Y\otimes X \otimes Z ,
\quad 
\mathfrak{p}_{23}(X\otimes Y \otimes Z)= X\otimes Z \otimes Y , 
\\
\mathfrak{p}_{13}(X\otimes Y \otimes Z)= Z\otimes Y \otimes X 
\qquad \text{for}
\qquad X,Y,Z \in U_{q}(\widehat{sl_{2}}).
\end{multline}
Applying 
$\mathfrak{p}_{13} \circ \mathfrak{p}_{12}$ 
 to \eqref{YBE}, we obtain
\begin{align}
\Rc_{23}\Rcb_{12}\Rcb_{13}&=\Rcb_{13}\Rcb_{12}\Rc_{23}.
\label{YBE2}
\end{align}
Then we evaluate  \eqref{YBE2} under $\rho^{(1)}_{x} \otimes \pi_{y} \otimes 1$
 ($x,y \in {\mathbb C}^{\times}$), to get
\begin{align}
\Lc_{23}(y)\Lfb^{(a)}_{12}(xy^{-1})\Lcb_{13}(x)&=\Lcb_{13}(x)\Lfb^{(a)}_{12}(xy^{-1})\Lc_{23}(y).
\label{YBEL2}
\end{align}
Now we can show the relation \eqref{TQ-uni} for $a=1$ step by step as follows
\footnote{Here $ \mathrm{tr}_{1}= \mathrm{tr}_{W_{1}} \otimes 1 \otimes 1$, 
$ \mathrm{tr}_{2}= 1 \otimes  \mathrm{tr}  \otimes 1$, 
and the third component of the tensor product is in $U_{q}(\widehat{sl_{2}})$.
The parts which contribute to the trace $ \mathrm{tr}_{1}$ 
 are linear combinations of the terms of the from $\es_1^{n}\fs_1^{n}q^{\xi \hs_1}$, 
 $n \in \mathbb{Z}_{\ge 0}$, $\xi \in \mathbb{C}$.  
 They are invariant under the anti-involution \eqref{t-osc}:  $(\es_1^{n}\fs_1^{n}q^{\xi \hs_1})^{t}=\es_1^{n}\fs_1^{n}q^{\xi \hs_1}$.
 We will also use the invariance  $g^{t}=g$, $\check{\overline{\mathbf K}}^{(1)}(x)^{t}=\check{\overline{\mathbf K}}^{(1)}(x)$ and 
 $\overline{K}(x)^{t}=\overline{K}(x) $.}.
\begin{align}
&(q^{2}-q^{4}x^{2s}) {\mathcal Q}^{(a)} (q^{\frac{1}{s}}x) {\mathcal T}(x)
\nonumber \\
& =
(q^{2}-q^{4}x^{2s}) \mathrm{tr}_{1} 
\left(
\check{\overline{\mathbf K}}^{(1)}_{1}(x^{-1}q^{-\frac{1}{s}}) 
{\mathcal K}^{(1)}_{13}(xq^{\frac{1}{s}}) 
\right)
\mathrm{tr}_{2}
\left(
\overline{K}_{2}(x^{-1}) 
{\mathcal K}_{23}(x)
\right)
\nonumber \\
& =
(q^{2}-q^{4}x^{2s}) \mathrm{tr}_{12} 
\left(
\check{\overline{\mathbf K}}^{(1)}_{1}(x^{-1}q^{-\frac{1}{s}})^{t_{1}} 
{\mathcal K}^{(1)}_{13}(xq^{\frac{1}{s}})^{t_{1}}
\overline{K}_{2}(x^{-1}) 
{\mathcal K}_{23}(x)
\right)
\nonumber \\
& =
(q^{2}-q^{4}x^{2s}) \mathrm{tr}_{12} 
\left(
\check{\overline{\mathbf K}}^{(1)}_{1}(x^{-1}q^{-\frac{1}{s}})
\overline{K}_{2}(x^{-1}) 
{\mathcal K}^{(1)}_{13}(xq^{\frac{1}{s}})^{t_{1}}
{\mathcal K}_{23}(x)
\right)
\nonumber \\
& =
\mathrm{tr}_{12} 
\Bigl(
\check{\overline{\mathbf K}}^{(1)}_{1}(x^{-1}q^{-\frac{1}{s}})
\overline{K}_{2}(x^{-1}) 
\underbrace{g_{2}^{-1}\Lfcb^{(1)}_{12} (x^{-2}q^{-\frac{5}{s}})^{t_{2}} g_{2}\Lf^{(1)}_{12} (x^{-2}q^{-\frac{1}{s}})^{t_{2}}}_{\text{from \eqref{LcbL1s=c}}}
{\mathcal K}^{(1)}_{13}(xq^{\frac{1}{s}})^{t_{1}}
{\mathcal K}_{23}(x)
\Bigr) 
\nonumber \\
& =
\mathrm{tr}_{12} \Bigl(
\Bigl( \check{\overline{\mathbf K}}^{(1)}_{1}(x^{-1}q^{-\frac{1}{s}})
g_{2}\Lfcb^{(1)}_{12} (x^{-2}q^{-\frac{5}{s}}) g_{2}^{-1}
\overline{K}_{2}(x^{-1})^{t_{2}}
\Bigr)^{t_{2}}
\nonumber \\
& \qquad \qquad \cdot
\Bigl(
\underbrace{{\mathcal K}^{(1)}_{13}(xq^{\frac{1}{s}})}_{\text{apply \eqref{dressed-UK-Q}}}
\underbrace{\Lf^{(1)}_{12} (x^{-2}q^{-\frac{1}{s}})^{t_{1}t_{2}}}_{\text{apply \eqref{tt}}}
\underbrace{{\mathcal K}_{23}(x)}_{\text{apply \eqref{dressed-UKti}}}
\Bigr)^{t_{1}} \Bigr)
\nonumber \\
& =
\mathrm{tr}_{12} \Bigl(
\Bigl( \check{\overline{\mathbf K}}^{(1)}_{1}(x^{-1}q^{-\frac{1}{s}})
g_{2}\Lfcb^{(1)}_{12} (x^{-2}q^{-\frac{5}{s}}) g_{2}^{-1}
\overline{K}_{2}(x^{-1})
\Bigr)^{t_{2}} 
\nonumber \\
&
\qquad \qquad 
\cdot 
\Bigl(
{\mathcal L}^{(1)}_{13}(x^{-1}q^{-\frac{1}{s}}) \Kf_{1}^{(1)}(xq^{\frac{1}{s}})  \underbrace{ \overline{\mathcal L}^{(1)}_{13}(xq^{\frac{1}{s}})
\Lfb^{(1)}_{12} (x^{2}q^{\frac{1}{s}})
{\mathcal L}_{23}(x^{-1}) }_{\text{apply \eqref{YBEL2}}}   K_{2}(x) \overline{\mathcal L}_{23}(x)
\Bigr)^{t_{1}} \Bigr)
\nonumber \\
& =
\mathrm{tr}_{12} \Bigl(
\Bigl( \check{\overline{\mathbf K}}^{(1)}_{1}(x^{-1}q^{-\frac{1}{s}})
g_{2}\Lfcb^{(1)}_{12} (x^{-2}q^{-\frac{5}{s}}) g_{2}^{-1}
\overline{K}_{2}(x^{-1})
\Bigr)^{t_{2}}
\nonumber \\
&
\qquad \qquad 
\cdot 
\Bigl(
{\mathcal L}^{(1)}_{13}(x^{-1}q^{-\frac{1}{s}}) 
\underbrace{ \Kf_{1}^{(1)}(xq^{\frac{1}{s}})  {\mathcal L}_{23}(x^{-1})}_{\text{exchange}} 
 \Lfb^{(1)}_{12} (x^{2}q^{\frac{1}{s}}) 
\underbrace{ \overline{\mathcal L}^{(1)}_{13}(xq^{\frac{1}{s}})  K_{2}(x)}_{\text{exchange}} 
\overline{\mathcal L}_{23}(x)
\Bigr)^{t_{1}} \Bigr)
\nonumber \\
& =
\mathrm{tr}_{12} \Bigl(
\Bigl( \check{\overline{\mathbf K}}^{(1)}_{1}(x^{-1}q^{-\frac{1}{s}} )
g_{2}\Lfcb^{(1)}_{12} (x^{-2}q^{-\frac{5}{s}}) g_{2}^{-1}
\overline{K}_{2}(x^{-1})
\Bigr)^{t_{1}t_{2}}
\nonumber \\
&
\qquad \qquad 
\cdot 
\Bigl(
 {\mathcal L}^{(1)}_{13}(x^{-1}q^{-\frac{1}{s}})  {\mathcal L}_{23}(x^{-1}) 
   \Kf_{1}^{(1)}(xq^{\frac{1}{s}}) \Lfb^{(1)}_{12} (x^{2}q^{\frac{1}{s}}) 
 K_{2}(x) 
  \overline{\mathcal L}^{(1)}_{13}(xq^{\frac{1}{s}}) 
\overline{\mathcal L}_{23}(x)  
\Bigr) \Bigr)
\nonumber \\
& =
\mathrm{tr}_{12} \Bigl(
\Bigl(\G_{12}^{t_{1}t_{2}} \underbrace{\check{\overline{\mathbf K}}^{(1)}_{1}(x^{-1}q^{-\frac{1}{s}} )
g_{2}\Lfcb^{(1)}_{12} (x^{-2}q^{-\frac{5}{s}}) g_{2}^{-1}
\overline{K}_{2}(x^{-1})(\Gb_{12}^{-1})^{t_{1}t_{2}}
}_{\text{apply \eqref{GKbLbKbG}}} 
\Bigr)^{t_{1}t_{2}}
\nonumber \\
& \qquad \qquad 
\cdot 
\Bigl(
 \underbrace{ \G_{12}^{-1}{\mathcal L}^{(1)}_{13}(x^{-1}q^{-\frac{1}{s}})  {\mathcal L}_{23}(x^{-1}) \G_{12}}_{\text{apply \eqref{GL1LGuni}}} 
 \cdot 
  \underbrace{ \G_{12}^{-1} \Kf_{1}^{(1)}(xq^{\frac{1}{s}}) \Lfb^{(1)}_{12} (x^{2}q^{\frac{1}{s}}) 
 K_{2}(x) \Gb_{12}}_{\text{apply \eqref{GKLbKG}}}  
 \nonumber \\
& \qquad \qquad \cdot
 \underbrace{\Gb_{12}^{-1} \overline{\mathcal L}^{(1)}_{13}(xq^{\frac{1}{s}}) 
\overline{\mathcal L}_{23}(x)  \Gb_{12}}_{\text{apply \eqref{GLb1LGuni}}}
\Bigr) \Bigr)
\nonumber \\
& =
\mathrm{tr}_{12} \Bigl( 
\Bigl( 
\overline{\omega}^{(1)}_{1}(x)
q^{-(\frac{2s_{0}}{s}-\frac{1}{2})\hs_{1}} \Kfcb^{(1)}(x^{-1}q^{\frac{1}{s}})
 \otimes E_{11} \otimes 1
 \nonumber \\
 & + \overline{\omega}^{(1)}_{2}(x)
  q^{(\frac{2s_{0}}{s}-\frac{1}{2})\hs_{1}} \Kfcb^{(1)}(x^{-1}q^{-\frac{3}{s}})
 \otimes E_{22} \otimes 1
 +\overline{\omega}^{(1)}_{12}(x)
  q^{-\frac{3}{2}\hs_{1}} \Kfcb^{(1)}(x^{-1}q^{-\frac{1}{s}})
 \fs_{1}
 \otimes E_{12} \otimes 1
 \Bigr)^{t_{1}t_{2}}
 \nonumber \\
 & \qquad \qquad \cdot \Bigl( 
  (q^{\frac{s_{0}-s_{1}}{2s}\hs_{1}} \otimes 1 \otimes 1) \Lc^{(1)}_{13}(x^{-1}q^{\frac{1}{s}})
(q^{-\frac{s_{0}-s_{1}}{2s}\hs_{1}} \otimes E_{11} \otimes q^{\frac{1}{2}h_{1}})
\nonumber \\
&\qquad \qquad  +
 (q^{-\frac{s_{0}-s_{1}}{2s}\hs_{1}} \otimes 1 \otimes 1) \Lc^{(1)}_{13}(x^{-1}q^{-\frac{3}{s}})
(q^{\frac{s_{0}-s_{1}}{2s}\hs_{1}} \otimes E_{22} \otimes q^{-\frac{1}{2}h_{1}})
\nonumber  \\
&
\qquad \qquad + \lambda x^{-s_{0}} {\mathcal F}_{13}(x^{-1}) (1 \otimes E_{21} \otimes q^{\frac{1}{2}h_{1}})
 \Bigr)
\cdot \Bigl( 
\omega^{(1)}_{1}(x) q^{(\frac{2s_{0}}{s}-\frac{1}{2})\hs_{1}} \Kf^{(1)}(xq^{-\frac{1}{s}})
 \otimes E_{11} \otimes 1
\nonumber \\
&\qquad  \qquad  +
 \omega^{(1)}_{2}(x) q^{-(\frac{2s_{0}}{s}-\frac{1}{2})\hs_{1}} \Kf^{(1)}(xq^{\frac{3}{s}})
 \otimes E_{22} \otimes 1
 +
 \omega^{(1)}_{21}(x) q^{-\frac{1}{2}\hs_{1}} \Kf^{(1)}(xq^{\frac{1}{s}})
 \es_{1}
 \otimes E_{21} \otimes 1
 \Bigr) 
\nonumber \\
& \qquad  \qquad  \cdot \Bigl(
(q^{\frac{s_{1}-s_{0}}{2s}\hs_{1}} \otimes 1 \otimes 1) \Lcb^{(1)}_{13}(xq^{-\frac{1}{s}})
(q^{-\frac{s_{1}-s_{0}}{2s}\hs_{1}} \otimes E_{11} \otimes q^{\frac{1}{2}h_{1}})
\nonumber \\
&\qquad  \qquad  +
 (q^{-\frac{s_{1}-s_{0}}{2s}\hs_{1}} \otimes 1 \otimes 1) \Lcb^{(1)}_{13}(xq^{\frac{3}{s}})
(q^{\frac{s_{1}-s_{0}}{2s}\hs_{1}} \otimes E_{22} \otimes q^{-\frac{1}{2}h_{1}})
\nonumber \\
&\qquad  \qquad  +
 x^{-s_{1}} \overline{\mathcal F}_{13}(x)
(1 \otimes E_{21} \otimes q^{\frac{1}{2}h_{1}})
\Bigr)
\Bigr)
\nonumber \\
& =\mathrm{tr}_{12} \Bigl( 
\omega^{(1)}_{1}(x)\overline{\omega}^{(1)}_{1}(x)  \Kfcb_{1}^{(1)}(x^{-1}q^{\frac{1}{s}}) 
 \Lc^{(1)}_{13}(x^{-1}q^{\frac{1}{s}})  \Kf^{(1)}_{1}(xq^{-\frac{1}{s}}) 
\Lcb^{(1)}_{13}(xq^{-\frac{1}{s}})
(1 \otimes E_{11} \otimes q^{h_{1}})
\nonumber \\
& \qquad  \qquad +
\omega^{(1)}_{2}(x)\overline{\omega}^{(1)}_{2}(x)  \Kfcb_{1}^{(1)}(x^{-1}q^{-\frac{3}{s}}) 
 \Lc^{(1)}_{13}(x^{-1}q^{\frac{3}{s}})  \Kf^{(1)}_{1}(xq^{\frac{3}{s}}) 
\Lcb^{(1)}_{13}(xq^{\frac{3}{s}})
(1 \otimes E_{22} \otimes q^{-h_{1}})
\nonumber \\
& \qquad \qquad +((\dots ) \otimes E_{21} \otimes (\dots ))
\Bigr) 
\qquad \text{[by  images of \eqref{uniCar} and cyclicity of the trace]}
\nonumber \\
&=\omega^{(1)}_{1}(x)\overline{\omega}^{(1)}_{1}(x) {\mathcal Q}^{(1)} (xq^{-\frac{1}{s}})q^{h_{1}}+
\omega^{(1)}_{2}(x)\overline{\omega}^{(1)}_{2}(x) {\mathcal Q}^{(1)} (xq^{\frac{3}{s}})q^{-h_{1}}
\nonumber \\
& \qquad \qquad \text{[by $\mathrm{tr}E_{11}=\mathrm{tr}E_{22}=1$, $\mathrm{tr}E_{21}=0$]} ,
\end{align}
where $(\dots)$ are the parts which do not contribute to the trace. 
Applying the  map $\zeta \circ \sigma $ to \eqref{TQ-uni} for $a=1$, one can show 
 \eqref{TQ-uni} for $a=2$. 
\section*{Appendix C: Proof of the universal dressed reflection equation}
\label{ApC}
\addcontentsline{toc}{section}{Appendix C}
\def\theequation{C\arabic{equation}}
\setcounter{equation}{0}
In this section, we prove \eqref{RE-dressL-Qti} 
 based on a universal version of Sklyanin's method \cite{Skly}. 
Applying 
$ \mathfrak{p}_{23}$ and 
 $\mathfrak{p}_{23} \circ \mathfrak{p}_{13} \circ \mathfrak{p}_{12}$, respectively,  
 to \eqref{YBE}, we obtain
\begin{align}
%
\Rc_{13}\Rc_{12}\Rcb_{23}&=\Rcb_{23}\Rc_{12}\Rc_{13},
\label{YBE3}
\\
\Rcb_{12}\Rcb_{13}\Rcb_{23}&=\Rcb_{23}\Rcb_{13}\Rcb_{12}.
\label{YBE4}
\end{align}
Then we evaluate  \eqref{YBE} and \eqref{YBE3}-\eqref{YBE4} under $\rho^{(a)}_{x} \otimes \pi_{y} \otimes 1$
 ($a=1,2$; $x,y \in {\mathbb C}^{\times}$), to get
\begin{align}
\Lf^{(a)}_{12}(xy^{-1})\Lc^{(a)}_{13}(x)\Lc_{23}(y)&=\Lc_{23}(y)\Lc^{(a)}_{13}(x)\Lf^{(a)}_{12}(xy^{-1}), 
\label{YBEL}
\\
%
\Lc^{(a)}_{13}(x)\Lf^{(a)}_{12}(xy^{-1})
\Lcb_{23}(y)&=\Lcb_{23}(y)\Lf^{(a)}_{12}(xy^{-1})\Lc^{(a)}_{13}(x), 
\label{YBEL3}
\\
\Lcb_{23}(y)\Lcb^{(a)}_{13}(x)\Lfb_{12}(xy^{-1})&=\Lfb_{12}(xy^{-1})\Lcb^{(a)}_{13}(x)\Lcb_{23}(y) .
\label{YBEL4}
\end{align}
One can prove  \eqref{RE-dressL-Qti} step by step as follows:
\begin{align}
&\Lf^{(a)}_{12} \left(x^{-1} y \right)
\underbrace{{\mathcal K}^{(a)}_{13}(x)  }_{\text{apply } \eqref{dressed-UK-Q}} 
\overline{\Lf}^{(a)}_{12} \left( xy \right) 
\underbrace{{\mathcal K}_{23}(y) }_{\text{apply } \eqref{dressed-UKti}} =
\nonumber \\
&=\underbrace{\Lf^{(a)}_{12} \left(x^{-1} y \right){\mathcal L}^{(a)}_{13}(x^{-1})}_{\text{apply } \eqref{YBEL}}
 \Kf^{(a)}_{1}(x) 
\underbrace{\overline{\mathcal L}^{(a)}_{13}(x)
\overline{\Lf}^{(a)}_{12} \left( xy \right) 
{\mathcal L}_{23}(y^{-1})}_{\text{apply } \eqref{YBEL2}}
 K_{2}(y) \overline{\mathcal L}_{23}(y)
\nonumber \\
&={\mathcal L}_{23}(y^{-1}){\mathcal L}^{(a)}_{13}(x^{-1})
\Lf^{(a)}_{12} \left(x^{-1} y \right)
\underbrace{ {\mathcal L}_{23}(y^{-1})^{-1} \Kf^{(a)}_{1}(x) {\mathcal L}_{23}(y^{-1})  }_{=\Kf^{(a)}_{1}(x)}
\overline{\Lf}^{(a)}_{12} \left( xy \right) 
\underbrace{ \overline{\mathcal L}^{(a)}_{13}(x) K_{2}(y) }_{=K_{2}(y) \overline{\mathcal L}^{(a)}_{13}(x)}
\overline{\mathcal L}_{23}(y)
\nonumber \\
&={\mathcal L}_{23}(y^{-1}){\mathcal L}^{(a)}_{13}(x^{-1})
\underbrace{
\Lf^{(a)}_{12} \left(x^{-1} y \right)
 \Kf^{(a)}_{1}(x) 
  \overline{\Lf}^{(a)}_{12} \left( xy \right) K_{2}(y)
 }_{\text{apply } \eqref{refeqlim1st}}
   \overline{\mathcal L}^{(a)}_{13}(x)
 \overline{\mathcal L}_{23}(y)
\nonumber \\
&={\mathcal L}_{23}(y^{-1})
\underbrace{ {\mathcal L}^{(a)}_{13}(x^{-1}) K_{2}(y) }_{=K_{2}(y) {\mathcal L}^{(a)}_{13}(x^{-1}) }
 \Lf^{(a)}_{12} (x^{-1}y^{-1})  \Kf^{(a)}_{1}(x) \overline{\Lf}^{(a)}_{12} (xy^{-1})
   \overline{\mathcal L}^{(a)}_{13}(x)
 \overline{\mathcal L}_{23}(y)
\nonumber \\
&=\underbrace{{\mathcal K}_{23}(y)}_{\text{from } \eqref{dressed-UKti}}
\underbrace{
\overline{\mathcal L}_{23}(y)^{-1}
{\mathcal L}^{(a)}_{13}(x^{-1}) \Lf^{(a)}_{12} (x^{-1}y^{-1}) 
 }_{\text{apply } \eqref{YBEL3}}
  \Kf^{(a)}_{1}(x) \overline{\Lf}^{(a)}_{12} (xy^{-1})
   \overline{\mathcal L}^{(a)}_{13}(x)
 \overline{\mathcal L}_{23}(y) 
\nonumber \\
&={\mathcal K}_{23}(y)
 \Lf^{(a)}_{12} (x^{-1}y^{-1}) {\mathcal L}^{(a)}_{13}(x^{-1}) 
\underbrace{  \overline{\mathcal L}_{23}(y)^{-1} \Kf^{(a)}_{1}(x) }_{= \Kf^{(a)}_{1}(x) \overline{\mathcal L}_{23}(y)^{-1} }
 \overline{\Lf}^{(a)}_{12} (xy^{-1})
   \overline{\mathcal L}^{(a)}_{13}(x)
 \overline{\mathcal L}_{23}(y) 
\nonumber \\
&={\mathcal K}_{23}(y)
 \Lf^{(a)}_{12} (x^{-1}y^{-1}) 
\underbrace{  {\mathcal K}^{(a)}_{13}(x)}_{\text{from } \eqref{dressed-UK-Q}} 
\underbrace{  \overline{\mathcal L}^{(a)}_{13}(x)^{-1}  
 \overline{\mathcal L}_{23}(y)^{-1}
 \overline{\Lf}^{(a)}_{12} (xy^{-1}) 
 }_{\text{apply } \eqref{YBEL4}} 
   \overline{\mathcal L}^{(a)}_{13}(x)
 \overline{\mathcal L}_{23}(y) 
 \nonumber \\
&={\mathcal K}_{23}(y)
 \Lf^{(a)}_{12} (x^{-1}y^{-1}) {\mathcal K}^{(a)}_{13}(x)
  \overline{\Lf}^{(a)}_{12} (xy^{-1})
\underbrace{   \overline{\mathcal L}_{23}(y)^{-1} \overline{\mathcal L}^{(a)}_{13}(x)^{-1}  
   \overline{\mathcal L}^{(a)}_{13}(x)
 \overline{\mathcal L}_{23}(y) 
  }_{=1} 
\nonumber \\
&={\mathcal K}_{23}(y)
 \Lf^{(a)}_{12} \left(x^{-1}y^{-1} \right) {\mathcal K}^{(a)}_{13}(x) 
 \overline{\Lf}^{(a)}_{12} \left(x y^{-1} \right),
 \quad a=1,2.
\label{proofDRE}
\end{align}
One can also prove
\footnote{We also remark that a universal dressed reflection equation in \cite{BT18} (eq. (G.7) in \cite{BT18}) 
can be proven similarly (evaluate  \eqref{YBE} and \eqref{YBE2}, \eqref{YBE3} and \eqref{YBE4} under $\mathsf{ev}_{x} \otimes \pi_{y} \otimes 1$
$x,y \in {\mathbb C}^{\times}$). In fact, \eqref{RE-dressL-Qti} (for $a=1$) is a limit of eq. (G.7) in \cite{BT18} 
($q^{-\mu} \to 0$, in the notation of  \cite{BT18}), and 
\eqref{RE-dressLti} is the image of eq. (G.7) in \cite{BT18}
 under $\pi \otimes 1 \otimes 1$.}
 \eqref{RE-dressLti} similarly 
(evaluate  \eqref{YBE}, \eqref{YBE2}, \eqref{YBE3} and \eqref{YBE4} under $\pi_{x} \otimes \pi_{y} \otimes 1$
$x,y \in {\mathbb C}^{\times}$). 
\section*{Appendix D: Generic unitarity relations of R-operators}
\label{ApD}
\addcontentsline{toc}{section}{Appendix D}
\def\theequation{D\arabic{equation}}
\setcounter{equation}{0}
It is known that the R-matrices (\eqref{Rmat1} and \eqref{Rmat2}) of the 6-vertex model satisfy the unitarity relation
\begin{align}
R(x) \overline{R}(x)=\overline{R}(x) R(x) =(q^{2}+q^{-2}-x^{s}-x^{-s})\mathbb{I} \otimes \mathbb{I},
\end{align}
where $\mathbb{I}$ is the $2 \times 2$ unit matrix. 
Here we reconsider this type of relations in general situation. 
We define the  generic R-operators  by 
\begin{align}
\Rf(xy^{-1})=\Rf(x,y) &= (\mathsf{ev}_{x} \otimes \mathsf{ev}_{y}) \Rc, 
\qquad 
\overline{\Rf}(xy^{-1})=\overline{\Rf}(x,y) = (\mathsf{ev}_{x} \otimes \mathsf{ev}_{y}) \overline{\Rc}, 
\end{align}
where $x,y \in \mathbb{C}^{\times}$. 
For  any finite dimensional irreducible representations $\mathcal{\chi}_{1},\mathcal{\chi}_{2}$ of $U_{q}(sl_{2})$, 
we set 
\begin{align}
R^{\chi_{1},\chi_{2}}(x) &= (\chi_{1} \otimes \chi_{2}) \Rf(x), 
\qquad 
\overline{R}^{\chi_{1},\chi_{2}}(x) = (\chi_{1} \otimes \chi_{2}) \overline{\Rf}(x).
\end{align}
Then the following relation holds \cite{FR92}
\begin{align}
R^{\chi_{1},\chi_{2}}(x)  \overline{R}^{\chi_{1},\chi_{2}}(x) 
=\overline{R}^{\chi_{1},\chi_{2}}(x)  R^{\chi_{1},\chi_{2}}(x)  
=S(x) (\mathbb{I}_{1} \otimes \mathbb{I}_{2}),
\label{uni-fin}
\end{align}
where $S(x)$ is a scalar function
\footnote{$S(x)=1$ in the normalization of the R-matrices in \cite{FR92}. But this is not 
the case with our R-matrices.}
 on $x$, and $\mathbb{I}_{1}$ and $\mathbb{I}_{2}$ are unit matrices.
This implies the following generic unitarity relation:
\begin{align}
\Rf(x) \overline{\Rf}(x) =\overline{\Rf}(x) \Rf(x) =\mathbf{C}(x), 
\label{uni-gen}
\end{align}
where $\mathbf{C}(x)$ is  central on $U_{q}(sl_{2}) \otimes U_{q}(sl_{2})$. 
 Derivation of \eqref{uni-gen} from \eqref{uni-fin} is given as follows \cite{Jimbo19}.
Let $\mathfrak{g}$ be a finite dimensional Lie algebra. 
The following proposition is well known. 
\vspace{5pt}
\\
{\bf Proposition} 
{\it 
(Page 71, Proposition 5.11 in \cite{Ja96})
Let  $u \in U_{q}(\mathfrak{g})$. If $u$ annihilates all finite dimensional irreducible
\footnote{The word `irreducible' is not explicitly written in the corresponding proposition in \cite{Ja96}. However this is not matter since 
any finite dimensional representation of $ U_{q}(\mathfrak{g})$ is 
completely reducible.} $U_{q}(\mathfrak{g})$-modules, then 
$u=0$.
}
\vspace{5pt}
\\
Then one can show the following. 
\vspace{5pt}
\\
{\bf Corollary} 
{\it If $ u \in U_{q}(\mathfrak{g})$ is scalar on all finite dimensional irreducible
$U_{q}(\mathfrak{g})$-modules, then 
$u$ belongs to the center of $U_{q}(\mathfrak{g})$. }
\vspace{5pt}
\\
{\em Proof.} 
Take any element 
$a \in U_{q}(\mathfrak{g})$ and apply Proposition to $[u,a]$. 
Then one finds $[u,a]=0$. This means that $u$ is central since $a$ is arbitrarily. 
$\square$ 

Let us regard \eqref{uni-fin} as a matrix with respect to the second component of the tensor product. 
The matrix elements of this matrix are scalar (or 0) for any finite dimensional irreducible representation $\chi_{1}$
 of $U_{q}(sl_{2})$. Then Corollary  suggests \eqref{uni-gen}. 

\end{document}